\documentclass[twocolumn]{aastex62}
\usepackage{color}

\renewcommand{\H}{\mathrm{H}}
\newcommand{\e}{\mathrm{e}}
\newcommand{\B}{\mathrm{B}}

\newcommand{\ds}{\displaystyle}

\begin{document}
\title{Theoretical Model of Hydrogen Line Emission from Accreting Gas Giants}
\author[0000-0003-0568-9225]{Yuhiko Aoyama}
\affiliation{Department of Earth and Planetary Science, Graduate School of Science, The University of Tokyo, 7-3-1 Hongo, Bunkyo-ku, Tokyo 113-0033, Japan}
\email{yaoyama@eps.s.u-tokyo.ac.jp}
\author[0000-0002-5658-5971]{Masahiro Ikoma}
\affiliation{Department of Earth and Planetary Science, Graduate School of Science, The University of Tokyo, 7-3-1 Hongo, Bunkyo-ku, Tokyo 113-0033, Japan}
\affiliation{Research Center for the Early Universe (RESCEU), Graduate School of Science, The University of Tokyo, 7-3-1 Hongo, Bunkyo-ku, Tokyo 113-0033, Japan}
\author[0000-0002-5964-1975]{Takayuki Tanigawa}
\affiliation{National Institute of Technology, Ichinoseki College, Takanashi, Hagisho, Ichinoseki-shi 021-8511, JapaBn}
\shorttitle{Hydrogen line emission from protoplanets}

\begin{abstract}
Progress in understanding of giant planet formation has been hampered by a lack of observational constraints to growing protoplanets. 
Recently, detection of an H$\alpha$-emission excess via direct imaging was reported for the protoplanet LkCa~15b orbiting the pre-main-sequence star LkCa~15. 
However, the physical mechanism for the H$\alpha$ emission is poorly understood.
According to recent high-resolution three-dimensional hydrodynamic simulations of the flow accreting onto protoplanets, the disk gas flows down almost vertically onto and collides with the surface of a circum-planetary disk at a super-sonic velocity and thus passes through a strong shockwave.
The shock-heated gas is hot enough to generate H$\alpha$ emission.
Here we develop a one-dimensional radiative hydrodynamic model of the flow after the shock by detailed calculations of chemical reactions and electron transitions in hydrogen atoms, and quantify the hydrogen line emission in the Lyman-, Balmer-, and Paschen-series from the accreting gas giant system. 
We then demonstrate that the H$\alpha$ intensity is strong enough to be detected with current observational technique. 
Comparing our theoretical H$\alpha$ intensity with the observed one from LkCa~15b, 
we constrain the protoplanet mass and the disk gas density. 
Observation of hydrogen line emission from protoplanets is highly encouraged to obtain direct constraints of accreting gas giants, which will be key in understanding the formation of gas giants. 
\end{abstract}


\section{Introduction}

The origins of the solar system and diverse extrasolar systems have yet to be revealed. 
In particular, the formation of gas giants would be a high-priority issue,  
because gas giants are so massive that they have had a dynamical influence on whole planetary systems.
Planets are formed in circum-stellar gas disks (or protoplanetary disks) \citep[e.g.][]{Hayashi1981}. 
A widespread idea, which is called the core accretion model, is that once a core grows to a critical mass via solid accretion, runaway gas accretion of the disk gas takes place and results in forming a massive envelope \citep[e.g.][]{Mizuno1980, Pollack+1996, Ikoma+2000}.
It is, however, still uncertain how and when they form.

Progress in understanding of gas giant formation is hampered by a lack of  direct observational constraints to growing protoplanets. 
The typical formation timescale of gas giants, which is constrained from the observationally inferred lifetime of protoplanetary disks, is at most 10~Myr \citep[e.g.][]{Hernandez+2008}. 
Although an increasing number of young exoplanets have been recently detected \citep[e.g., CI Tau b][]{Johns-Krull+2016},
most of gas giants detected so far are several billion years old (e.g., see exoplanet.eu),
Those aged gas giants hardly have memory of their formation processes \citep[e.g.][]{Marley+2007}.

A challenging issue would be to detect accreting gas giants. 
Recent observations have detected infra-red (IR) excess from the three young stars, LkCa15 \citep{Kraus&Ireland2012}, HD169142 \citep{Biller+2014, Reggiani+2014}, and HD100546 \citep{Quanz+2015}.
Those observed excess is interpreted as infra-red (IR) emission from accreting gas giants \citep{Zhu2015}. 
Hydrodynamic simulations of gas accretion onto protoplanets show that accreting gas giants are surrounded by circum-planetary disks (CPDs hereafter) \citep[e.g.][]{Miki1982,Tanigawa+2002}. 
Then, the CPD gas falls toward the gas giant, losing its angular momentum through spiral shock waves and turbulent dissipation.
Since the angular momentum loss leads to conversion from gravitational energy to thermal energy, the CPD gas is warmer than the original circum-stellar disk gas. 
According to theoretical modelling, CPDs are warm and geometrically large enough to generate detectable IR emission \citep{Zhu2015}.

Among those stars, in additional to IR, an excess of hydrogen Balmer-$\alpha$ line (H$\alpha$) emission was detected in the circum-stellar disk of the young star LkCa15 of age 2 Myr \citep{Sallum+2015}.
In the case of protostars, it is well known that accretion shock near protostars brings about hydrogen line emission \citep{Lynden-Bell&Pringle1974}.
Likewise, the H$\alpha$ excess detected for LkCa15 is expected to arise from a shock-heated, accreting gas giant.

Theoretical models of stellar accretion developed so far, however, cannot be applied directly to planetary accretion.
In general, H$\alpha$ line emission occurs from hot hydrogen of tens of thousands kelvin, which is thought to be reached by accretion-shock heating.
In the case of stellar accretion, the strong magnetic field is thought to make a gap between the star and the circum-stellar disk.
Then, the accreting gas falls from the disk edge to the stellar surface, resulting in strong shocks \citep{Uchida&Shibata1984, Konigl1991}. 
The amount of energy generated by the accretion flow (i.e., released gravitational energy) depends on the flow structure.
The flow structure around planets is markedly different from that around stars, basically because planets are rotating around central stars.
Thus, it is necessary to develop a new model in order to explore whether accreting gas giants yield strong, observable H$\alpha$ emission.

Recent high-resolution three-dimensional hydrodynamical simulations of accretion flow onto protoplanets revealed that 
the flow enters the Roche lobe (or the Hill sphere) not through the Lagrange points in the midplane but from high altitudes \citep{Tanigawa+2012}.
Then, the vertically accreting flow hits the surface of the CPD.
Because the flow velocity is nearly free-fall velocity, which is much higher than the local sound speed, strong shock occurs at the CPD surface. 
In the extreme case of strong shock, the gas temperature reaches tens of thousands kelvin just behind the shock front, as shown later in this paper. 
In such high temperature regions, hydrogen line emission occurs.
From their 3D radiative hydrodynamical simulations, 
\citet{Marleau+2017} and \citet{Szulagyi&Mordasini2017} pointed out the presence of hot regions around accreting gas giants that could be the source of the observed H$\alpha$ line emission. 
However, they never quantified hydrogen line emission from those hot regions, 
because those regions, which are much thinner than the CPD thickness, hardly affect the CPD structure.

Radiative continuum emission from shock-heated gas was investigated so far for some other astronomical objects and events, 
which include white dwarf accretion \citep{Frank+1983}, protostar accretion \citep{Calvet+Gullbring1998, Lamzin1998}, 
the interstellar medium \citep{HM79,HM89,MacLow+Shull1986,Shapiro+Kang1987}, and chondrule formation in protoplanetary disks \citep{Iida+2001}. 
However, there is no detailed research focusing on hydrogen line emission from highly shock-heated gas, which is of interest in this study.
In the case of protostellar accretion, \citet{Calvet+Gullbring1998} investigated the shock heating and atomic line emission at the protostellar photosphere. 
They, however, assumed weak shock (or C-type shock), because of the magnetic effect, in contrast to strong shock which occurs in our problem.
\citet{Lamzin1998} also investigated the recombination lines emitted from the ionized atoms, 
which came not directly from the postshock gas but from the heated photosphere.
In the case of the interstellar shock, although \citet{HM79} investigated the hydrogen line emission, their estimation was simply based on optical depth and the escape probability approximation. 
Namely, they neglected the absorptive excitation and underestimated the excitation degree in optically thin regions.
While hydrogen level population certainly has little influence on the total luminosity from and cooling in the postshock regions, 
considering it is essential for estimation of each line luminosity.
Hence, in order to interpret the H$\alpha$ observation, one must consider transitions between energy levels in hydrogen in more detail. 

The purpose of this study is to quantify the hydrogen line emission at the surface of the CPD around an accreting gas giant. 
To that end, we investigate the hydrodynamic, thermochemical, and radiative properties of the vertically accreting flow after the passage of the shock front by performing 1D hydrodynamic simulations with detailed calculations of hydrogen level population. 
The details of the theoretical model and numerical method are presented in section~\ref{TM}.
Then, we show results of numerical simulations in section~\ref{R}, where we estimate the intensities of hydrogen line emission in the Lyman-, Balmer-, and Paschen-series. 
In section~\ref{D}, we demonstrate that we can obtain constraints to the mass of the accreting gas giant and the density of the surrounding disk gas from observed H$\alpha$ emission, by comparing between the theoretically estimated and measured H$\alpha$ luminosity for LkCa15 as an example. We also discuss the validity of our assumptions and future studies.
Finally, we summarize and conclude this study in section~\ref{CS}.

\section{Theoretical Model}
\label{TM}

\begin{figure*}[htb]
	\plotone{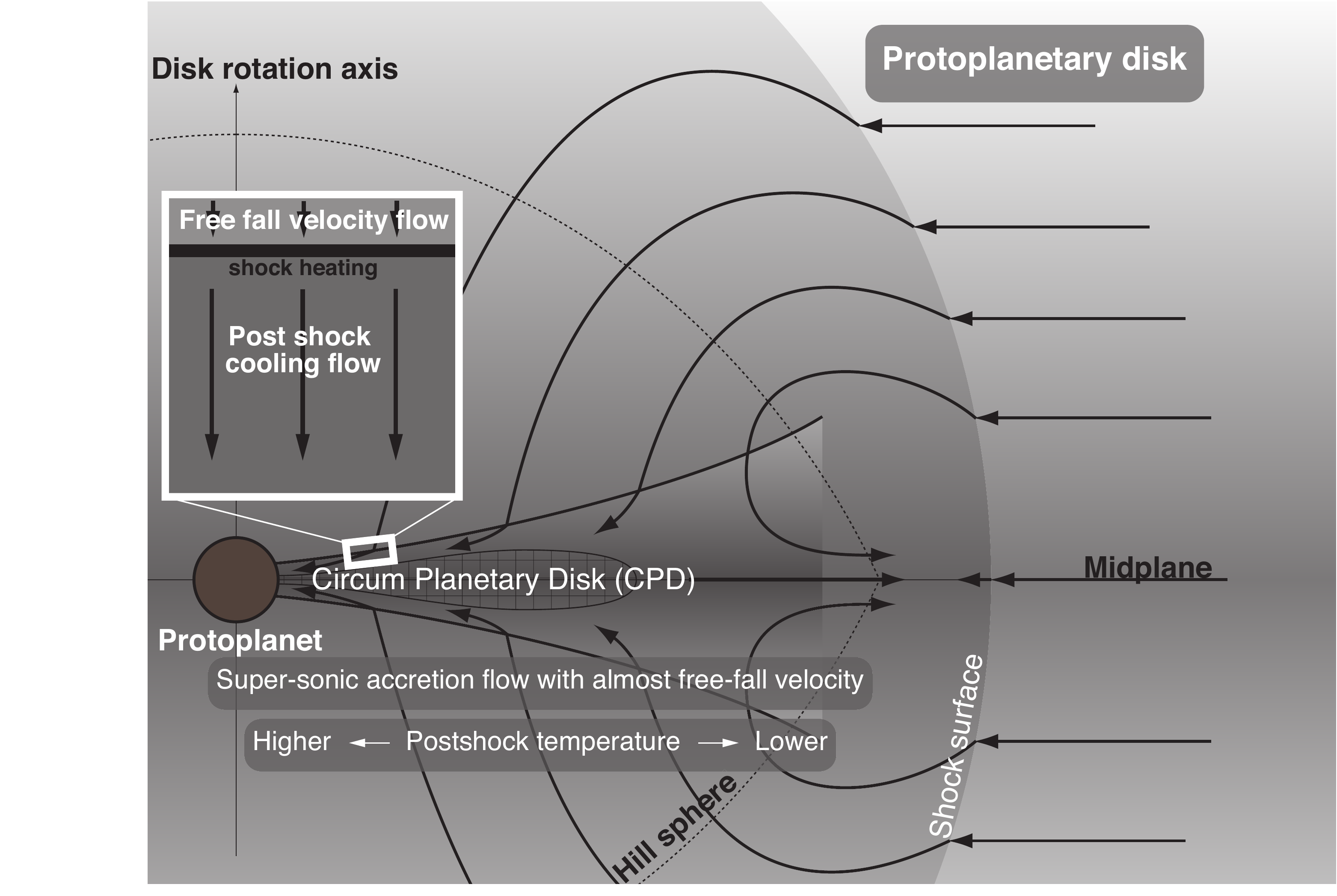}
	\caption{
	Schematic illustration of the circum-protoplanetary environment that we suppose in this study \citep[edge-on view, modified from][]{Tanigawa+2012}. 
	The arrows indicate streamlines of the accretion flow.
	The inset left above shows an enlarged illustration of the vicinity of the shock front for which we develop a theoretical model here.
	}
	\label{fig:schematic}
\end{figure*}

As described in Introduction, based on recent 3D simulations \citep[e.g.,][]{Tanigawa+2012}, we consider the situation in which the gas from the circum-stellar disk (CSD) flows almost vertically onto the circum-planetary disk (CPD) nearly at the free-fall velocity (see Fig.~\ref{fig:schematic} for a schematic illustration).
This type of flow is achieved when planet mass is large enough for the accreting gas to form a circum-planetary disk \citep[][]{Tanigawa+2002}.
Since the free-fall velocity is higher than the local sound velocity, 
shockwave is formed at the CPD surface. 
When passing through the CPD surface (i.e., the shock front), 
the gas is heated up to tens of thousands of kelvins, which is high enough to dissociate hydrogen molecules and ionize hydrogen atoms, producing free electrons. 
Then, the electrons collide with and excite hydrogens. 
After that, de-excitation of the excited hydrogen results in line emission and cooling. 

Thus, to calculate the intensities of hydrogen line emission from the CPD surface, we simulate chemical reactions, excitation/de-excitation of hydrogen atoms, and radiative cooling simultaneously with simulating hydrodynamics of the postshock gas. 
Here we describe our theoretical model that simulates  the hydrodynamic and thermal properties of the postshock gas.
All the physical processes and associated references are summarized in Table~\ref{tab:processes}.
\begin{table}[htbp]
\begin{center}
\caption{Physical Process list}	
	\begin{tabular}{l l }
		Physical Process & Reference \\ \hline \hline
		1D hydrodynamics & \citet{Shapiro+Kang1987}\\
		Chemical reaction & \citet{Iida+2001} \\
		Radiative transfer & \citet{Chandrasekhar1960}\\ \hline
		\multicolumn{2}{c}{Hydrogen Electron Transitions}\\
		Collisional transition & \citet{Vriens+Smeets1980}\\
		Spontaneous de-excitation &  \citet{Vriens+Smeets1980}\\
		Spontaneous recombination & \citet{Johnson1972}\\
		Photon induced transition & \citet{Castor2004}\\
		Photon absorptive ionization & \citet{Shu1991}\\ \hline
		\multicolumn{2}{c}{Cooling and Heating Processes}\\
		Hydrogen molecule dissociation & \citet{Blanksby+Ellison2003}\\
		Hydrogen atomic transition & \citet{Vriens+Smeets1980}\\
		Molecular lines & \citet{Iida+2001}\\ \hline		
	\end{tabular}
	\label{tab:processes}
\end{center}
\end{table}

\subsection{Key Assumptions}
We assume the shock heating as transient. 
Namely, the shockwave is jump type and regarded as an infinitely thin adiabatic layer, which is called a shock front. 
This approximation is valid, because the Mach number of the flow of interest is much larger than unity at the shock front ($\gtrsim$ 30). 
Also, the magnetic effect, which tends to reduce shock heating, can be ignored, because the preshock gas is too cold to ionize in gas giant forming regions which are usually far from host stars \citep[e.g.,][]{Ikoma+2000}.
Thus, in this study, without observing the interior of the shock front, we investigate the hydrodynamical and thermochemical properties of the flow only after the passage of the shock front.

We consider one-dimensional, plane parallel, hydrodynamically steady flow (see the inset of Fig.~\ref{fig:schematic}). 
This is valid because the thickness of the postshock region is much smaller than the CPD thickness. 
Note that the shock front is located a few scale-heights far from the CPD midplane, and thus the postshock flow is never affected by the CPD.
We follow the temporal change in properties of the gas flow with its motion, using the Lagrangian coordinates, and define the shock front as the origin. 

We assume that the gas is ideal and composed of the four elements H, He, C, and O and electrons.
The ideal approximation is valid because the temperature and density of the gas are sufficiently high and low, respectively.
We take the relative abundances of those four elements from \citet{Allen3rd}, namely $\rm H:He:C:O$ = $1:8.5\times 10^{-2}:3.3\times 10^{-4}:6.6\times 10^{-4}$.
We solve 160 chemical reactions that involve 33 gas species (see \S~\ref{TCR}) and 10 principal quantum numbers of hydrogen (see \S~\ref{TET} and \S~\ref{THI}), in addition to the ionized state. 
Inclusion of other elements such as N and S has little influence on the line emission intensities, because those are much less abundant than H and the regions where molecular cooling occurs are of little interest in this study, as shown later.
We consider the radiative transfer only of hydrogen lines and CO, OH, and H$_2$O molecular lines. 
Also, we assume that the electrons are the same in temperature as other gases, namely neglect the acceleration by electric and magnetic fields.

Finally, we neglect the presence of dust grains in the flow. This is a reasonable assumption, because the gas falling onto the inner CPD comes from high altitudes. 
It is thought that dust grains have already settled down gravitationally and exit in thin layers 
near the CSD mid-plane in planet formation stages. 
Thus, the high altitude gas hardly contains dust grains \citep{Goldreich+Ward1973}.
In addition, if any, small dust grains coupled with gas are quickly sublimated in the postshock gas because of high temperature ($\gg 10^4$K).
Although recondensation of silicate may occur when gas becomes cool enough, hydrogen line emission occurs at temperatures higher than the condensation temperature, which means such dust cooling is of little interest in this study.

\subsection{Hydrodynamics}
\subsubsection{Jump condition across the shock front} 
In the case of jump-type shock, mass, momentum, and energy are conserved across the shock front.
The relationship between the gas properties on both sides of the shock front is described, respectively, as follows \citep{Landau+Lifshitz1959}:
\begin{eqnarray}
\label{eq:mascon}
\rho_1 v_1 &=& \rho_0 v_0, \\
\label{eq:momcon}
\rho_1 v_1^2 + p_1 &=& \rho_0 v_0^2 +p_0, \\
\label{eq:enecon}
v_1\left(\frac{\rho_1 v_1^2}{2}+\frac{\gamma}{\gamma-1}p_1 \right)&=&v_0\left(\frac{\rho_0 v_0^2}{2}+\frac{\gamma}{\gamma-1}p_0 \right),
\end{eqnarray}
where $v_0$ $(v_1)$ is the preshock (postshock) velocity in the frame of the shock front, $\rho_0$ $(\rho_1)$ and $p_0$ $(p_1)$ are the density and pressure of the preshock (postshock) gas, respectively, and $\gamma$ is the specific heat ratio. 
Under the assumption of transient shock heating, all the abundances of chemical species and all the electron levels  remain unchanged across the shock front. 
Hence the specific heat ratio is assumed to be constant ($\gamma$ = 1.42, since
we assume that hydrogen is in its molecular form and the others are in their atomic forms at the shock front).
The postshock temperature, $T_1$, is given by the ideal equation of state as
\begin{equation}
\label{eq:EOS}
T_1=\frac{\mu \, p_1}{ k_\mathrm{B}\rho_1},
\end{equation}
where $k_\mathrm{B}$ is the Boltzmann constant and $\mu$ is the mean mass of the gas per particle.  
From the assumed molecular abundances, $\mu$ = $3.84\times 10^{-27}$~kg at the shock front.
In this paper, we have performed numerical simulations in the ranges of
$20~\mathrm{km/s} \leq v_0 \leq 100~\mathrm{km/s}$ and 
$10^{15} ~\mathrm{m^{-3}} \leq n_{\mathrm{H},0} \leq ~\mathrm{10^{20}\mathrm{m^{-3}}}$, where $n_{\mathrm{H},0}$ is the proton number density just before the shock. 
(Note that gas density becomes higher by a factor of $\sim$ 5 and $\sim$ 100, respectively, just after the shock and where hydrogen line emission occurs.)
Then, the gas temperature just after the shock, $T_1$, is up to $\sim 4\times 10^{5}$~K. 

\subsubsection{Postshock gas flow}
In the postshock region, mass and momentum are likewise conserved, but the adiabatic approximation (Eq.~[\ref{eq:enecon}]) is invalid. 
The postshock gas flow is described by the following three equations.
 \begin{eqnarray}
\rho v &=& \rho_1v_1,
\label{eq:massflow} \\
\rho v^2+p&=&\mathrm{A_1}\rho_1v_1^2,
\label{eq:momflow} \\
\frac{dE}{dt}&=&\left(\Gamma - \Lambda \right)+\left[ \frac{p+E}{\rho} \frac{d\rho}{dt}\right],
\label{eq:eneflow} 
\end{eqnarray}
where
$p$, $\rho$, and $v$ are the pressure, gas density, and fluid velocity in the frame of the shock front, respectively, 
\begin{equation}
	A_1 \equiv 1+ \frac{p_1 }{ \rho_1 v_1^2},
\end{equation}
$E$ is the internal thermal energy per unit volume, and $\Gamma$ and $\Lambda$ are the heating and cooling rates per unit volume, respectively.
The expressions of $\Gamma$ and $\Lambda$ are given in section~\ref{CHR}.
We integrate equations (\ref{eq:massflow})-(\ref{eq:eneflow}) numerically, following \citet{Shapiro+Kang1987}.

\subsection{Cooling and Heating Processes}
\label{CHR}
\subsubsection{Exothermic and endothermic chemical reactions}
Regarding the energy budget relevant to molecular chemical reactions, 
we consider only collisional dissociation and recombination of the major molecule $\H_2$ among the simulated reactions (see section~\ref{TCR}).
The corresponding rate of net energy change 
\begin{eqnarray}
\label{eq:h2d}
 (\Lambda - \Gamma)_{\H_2}  &=&-E_{\H_2} \frac{dn_{\H_2}}{dt} ,
\end{eqnarray}
where $E_{\H_2}$ is the binding energy of an $\H_2$ molecule \citep[$=435.998~\mathrm{kJ}$;][]{Blanksby+Ellison2003} and $n_\mathrm{H_2}$ is the number density of H$_2$ molecules.
In Eq.~(\ref{eq:h2d}), we have neglected the energy of rotation and vibration of H$_2$ molecules. 
The shock heating is strong enough to dissociate H$_2$ molecules completely. Also, the recombined H$_2$ is of little interest in this study. 
Hence, neglecting the rotational and vibrational energies barely affects our conclusion.

\subsubsection{Radiative cooling by molecules}
We take into account 
some major processes of radiative cooling by molecules, which include vibration of CO and  rotation of $\mathrm{H_2O}$ and OH.
The cooling rate due to CO vibrational emission is given as \citep{Iida+2001}
\begin{eqnarray}
\label{eq:coc}
 \Lambda_\mathrm{CO}  &=& 
n_\mathrm{CO}
\left[
\frac{1}
       {\ds \left(  \xi^\H_\mathrm{CO} n_\H +\xi^\mathrm{H_2}_\mathrm{CO} n_\mathrm{H_2}  \right) E_\mathrm{CO}} 
       + \frac{1}{\ds L_\mathrm{LTE}} 
\right]^{-1}
\end{eqnarray}
where 
$n_\mathrm{CO}$ and $n_\mathrm{H}$ are the number densities of CO molecules and isolated hydrogen atoms, respectively,
$E_\mathrm{CO}$ is the CO vibrational transition energy \citep[$\tilde{E}_\mathrm{CO} \equiv E_\mathrm{CO} / k_\B=3080\mathrm{K}$;
][]{Millikan+White1963},
 and $\xi^\H_\mathrm{CO}$ and $\xi^\mathrm{H_2}_\mathrm{CO}$ are the transition rates from the ground level to the first excited level $(v=1)$ by collision with H atoms and $\mathrm{H_2}$ molecules, respectively.
 In~Eq. (\ref{eq:coc}), we have neglected collisional excitation by minor gas species other than H and $\H_2$.
The above transition rates are given as \citep{HM89} 
\begin{eqnarray}
\xi^\H_\mathrm{CO}&=&3.0\times10^{-18} T^{0.5} \nonumber\\
&&\exp{\left[ -\left( \frac{C_1}{T} \right)^{3.43}-\left( \frac{\tilde{E}_\mathrm{CO}}{T} \right)\right]}\mathrm{m^3~s^{-1}},
\end{eqnarray}
and
\begin{eqnarray}
\xi^\mathrm{H_2}_\mathrm{CO}&=&4.3\times10^{-20} T\nonumber\\
&&\exp{\left[ -\left( \frac{C_2}{T} \right)^{0.333}-\left( \frac{\tilde{E}_\mathrm{CO}}{T} \right)\right]}\mathrm{m^3~s^{-1}},
\end{eqnarray}
where $T$ is the temperature in Kelvin, $C_1$ = $2.0 \times 10^3$~K, and $C_2$ = $3.14 \times 10^5$~K.
Also, $L_\mathrm{LTE}$ is the thermal emission per CO molecule whose level population is in the local thermodynamic equilibrium (LTE) and is given as \citep{Neufeld+Kaufman1993}
\begin{eqnarray}
L_\mathrm{LTE} = 1.0\times10^{-18} \exp{\left( -\frac{\tilde{E}_\mathrm{CO}}{T} \right)} \,\mathrm{J~s^{-1}}.
\end{eqnarray}

The cooling rate due to rotational transition of molecules $j$ with dipole moments ($j$ = H$_2$O  and OH) is expressed as \citep{HM79}
\begin{eqnarray}
\label{eq:rotrad}
\Lambda _{\mathrm{rot},j}=\frac{n_j (n_\H-n_{\H_2})\sigma v_\mathrm{th} k_\B T}{\ds 1+ ( n_\H  n^{-1}_\mathrm{cr}) \left[ 1+\ N_j  (A_j N_{1/2})^{-1} \right]}
\end{eqnarray}
where $v_\mathrm{th}$ is the thermal velocity of gas particles defined by $v_\mathrm{th}=\sqrt{8k_\B T  (\pi \mu)^{-1}}$, 
$N_j$ is the column density of species $j$ integrated from the shock front which is given by
$N_j$ = $\int _0^t n_j v dt$,
$\sigma$ is the total rotational de-excitation cross section of the molecules, 
$n_\mathrm{cr}$ is the critical number density above which the collisional deactivation overwhelms the spontaneous decay for the levels at which the former dominates cooling,
$N_{1/2}$ is the column density with which the cooling rate is half of that in the optically thin limit, and $A_j$ is the dipole moment. 
The values of the parameters used in Eq.~(\ref{eq:rotrad}) are given by \citet{HM79,HM89}.
This cooling rate, which is derived based on photon escape probability, is valid, regardless of optical thickness.

\subsubsection{Cooling due to collisional de-excitation}
\label{TCE}
Because of high temperature ($\gtrsim$ $1 \times 10^4$~K), 
a great number of free electrons are produced after the shock front. 
Those electrons collide with atoms and molecules and excite the atomic and molecular electron levels. 
Subsequent de-excitations result in radiative emission and make great contribution to cooling. 
In this study, regarding the collisional de-excitation, we take only the contribution of atomic hydrogen (i.e., isolated hydrogen atoms) into account, because the others are minor. 
Although the energy is removed eventually via radiation, the decrease in kinetic energy, which leads to reducing temperature, is due directly to collisional excitation and ionization. 
Thus, the cooling rate is given by
\begin{eqnarray}
\label{eq:ecc}
\Lambda_\mathrm{col} &=& -
n_\e \sum^{\mathfrak{N}}_{j=1} \sum^{\mathfrak{N}}_{i=j+1} 
\left[ 
\left( 	K_{ij}^\downarrow H_i - K_{ji}^\uparrow H_j \right)E_{ji} 
\right.
\nonumber\\
&+& \left. \left( K_{+ j}^\downarrow n_\e H_{+} - K_{j+}^\uparrow H_j \right)E_{j+} \right], 
\end{eqnarray}
where $\mathfrak{N}$ is the maximum principal quantum number taken into account (i.e., $\mathfrak{N}$ = 10 in this study), 
$H_i$ is the number density of the isolated hydrogen atoms whose principal quantum number is $i$ (i.e., $i$th level atomic hydrogen), 
$n_\e$ and $H_+$ are the number densities of the free electrons and ionized hydrogen (or hydrogen ions), respectively, 
$K_{ji}^\uparrow$ and $K_{ij}^\downarrow$ are the collisional excitation and de-excitation coefficients, respectively, for transition between the lower level $j$ and the upper level $i$, 
$E_{ji}$ ($> 0$) is the energy difference between the $j$th and $i$th levels in atomic hydrogen, and the subscript + represents the ionized state.
These transition coefficients are given in \citet{Vriens+Smeets1980}.
In Eq. (\ref{eq:ecc}), the first term in the square bracket is the effective cooling rate for collisional excitation from the $j$th to $i$th levels and
the second term is that for ionization from the $j$th level. 
When the de-excitation or recombination is dominant, each term in equation~(\ref{eq:ecc}) becomes negative, which means that heating occurs.

\subsection{Chemical Reaction}
\label{TCR}
As described in section~\ref{CHR}, to follow the cooling and heating processes, we have to know, at least, the abundances of $\H_2$, H, CO, OH, and $\e^-$. 
To simulate the temporal change in their abundances, 
we adopt the tuned chemical reaction system developed by \citet{Iida+2001} who also addresses the chemical processes in postshock gas in a protoplanetary disk. 
In this study we consider 33 chemical species composed of H, He, C, and O, and 160 chemical reactions, all of which are listed in Tables~1-4 of \citet{Iida+2001}.
In contrast to \citet{Iida+2001} being interested in molecular line cooling, we consider the excitation levels of hydrogen atoms before ionization and after recombination by Eq.~(\ref{eq:dion}), because we are interested in hydrogen line cooling (see section~\ref{THI}).
Also, \citet{Iida+2001} focuses on the fate of silicate dust grains and thus considers Si-bearing species and relevant reactions,
whereas we do not take them into account because we assume dust-free gas and can ignore line cooling by minor molecules.
We use the numerical package DLSODE in ODEPACK \citep{LSODE} in order to integrate the temporal change of chemical species, numerically.
This numerical method is also used for electron transitions (i.e. Eqs.~[\ref{eq:dele}] and [\ref{eq:dion}]).

\subsection{Bound-Bound Transition}
\label{TET}
The number density of hydrogen atoms of principal quantum number $i$, $H_i$, changes with time, because of excitation/de-excitation, ionization, and recombination by collision, photon absorption, induced photon emission, and spontaneous photon emission.
The temporal change in $H_i$ by transitions from and to the $j$th level due to collision with electrons, $T^\mathrm{C}_{ji}$, is given as
\begin{eqnarray}
\label{eq:tcol}
T^\mathrm{C}_{ji}&=&(K_{ji}^\uparrow H_{j} - K_{ij}^\downarrow H_{i} )n_\mathrm{e}.
\end{eqnarray}
The change in $H_i$ by spontaneous photon radiation is given as
\begin{equation}
\label{eq:tspn}
 T^\mathrm{A}_{ij} = A^{\mathrm{spn}}_{ij} H_i ,
\end{equation}
where $A^\mathrm{spn}_{ij}$ is the Einstein coefficient for spontaneous transition from the $i$th to $j$th levels.
The change in $H_i$ by photon absorption and induced radiation is given as \citep[e.g.][]{Castor2004}
\begin{eqnarray}
\label{eq:tabs}
T^\mathrm{B}_{ji}&=&\int_{0}^{\infty} u_{ij} {\left( B_{ji}^{\mathrm{abs}} H_j - B_{ij}^{\mathrm{ind}} H_i  \right) D } \, d\nu,
\end{eqnarray}
where $u_{ij}$ is the spectral energy density (i.e., energy flow per unit volume per unit wavelength interval) yielded by transition from the $i$th to $j$th levels and functions of the frequency $\nu$ (see section~\ref{Rad}),
 $B^\mathrm{abs}_{ji}$ and $B^\mathrm{ind}_{ij}$ are the Einstein coefficients for the absorptive and induced transitions between the $i$th and $j$th levels, respectively, 
and 
$D (\nu)$ is the spectral broadening function for which 
we use the approximated Voigt function derived by \cite{Humlicek1982} and consider the Doppler, natural, and pressure broadenings \citep[see e.g.][]{Castor2004}.

Then, the total temporal change in the number of the $i$th level hydrogen is given as
\begin{eqnarray}
\label{eq:dele}&&
\begin{array}{l l l}
\ds\frac{dH_{i}}{dt}&=&\ds \sum^{i-1}_{j=1} \left(T^\mathrm{C}_{ji} +T^\mathrm{B}_{ji} -T^\mathrm{A}_{ij} \right)  \\
&+&\ds \sum^{\mathfrak{N}}_{j=i+1} \left( - T^\mathrm{C}_{ji} -T^\mathrm{B}_{ji} +T^\mathrm{A}_{ij} \right)\\
&+&T_{+ i},
\end{array}
\end{eqnarray}
where $T_{+ i}$ is the transition rate from the ionized to $i$th level state of hydrogen (see Eq. [\ref{eq:dion}]).

\subsection{Radiative Transfer}
\label{Rad}
Radiative transfer in the postshock flow plays an essential role in determining the intensity of the hydrogen line emission from the shock surface.
In addition, absorption of radiation with energy equal to the difference between energy levels (i.e., photon resonant absorption)
changes the hydrogen excitation degree and, in some case, affects the temperature profile after the shock.
We perform two-stream integration of hydrogen line emission, namely the same and opposite directions relative to the gas flow.
Assuming the spectral energy density per unit angle as a quadratic function of $\sin{\theta}$ ($\equiv \mu)$, where $\theta$ is the angle measured from the direction of outward photon flow, 
we can analytically integrate this quadratic function and get the spectral energy density and flux as follows:
\begin{eqnarray}
u_{ij} 
=\frac{2\pi}{c}\left[ \frac{4}{3}I_{ij}(0) + \frac{1}{3}I_{ij}(1) + \frac{1}{3}I_{ij}(-1) \right] \\
F^\mathrm{u}_{ij}=\pi \left[ \frac{1}{2}I_{ij}(0) + \frac{7}{12}I_{ij}(1) + \frac{1}{12} I_{ij}(-1) \right]\\
F^\mathrm{d}_{ij}=\pi \left[ \frac{1}{2}I_{ij}(0) + \frac{1}{12}I_{ij}(1) + \frac{7}{12} I_{ij}(-1) \right]
\end{eqnarray}
where $F^\mathrm{u}_{ij}$, $F^\mathrm{d}_{ij}$, and $I_{ij}(\mu)$ are the upward energy flux, downward energy flux, and intensity of each line yielded by transition from the $i$th to $j$th levels and $\mu=1$, $-1$, and $0$ correspond to the directions that are the same as, opposite to, and perpendicular to the gas flow, respectively.
The spacial change in the intensity is given as \citep[e.g.][]{Chandrasekhar1960}
\begin{eqnarray}
\label{eq:RT}
\left(\frac{\partial}{\partial t} + \textbf{v} \cdot \nabla \right)
\left[ \mu I_{ij}(\mu) \right] = 
\nonumber \\
\left[ \frac{1}{c}(B_{ij}^\mathrm{ind}H_i - B_{ji}^\mathrm{abs} H_j) I_{ij} (\mu) +\frac{A^\mathrm{spn}H_i}{4\pi} \right] v h \nu
\end{eqnarray}
Under the assumptions of steady state and plane parallel structure,
the left-hand side in Eq.~(\ref{eq:RT}) must be zero.
Thus, the intensity for $\mu=0$ is given by
\begin{equation}
I(0)=\frac{c A_{ij}^\mathrm{spn}H_i} {4\pi (B_{ji}^\mathrm{abs}H_i- B_{ij}^\mathrm{ind} H_j ) }.
\end{equation}

To obtain steady flows in other directions, we 
integrate Eq.~(\ref{eq:RT})
with an explicit integration scheme,
unlike the case of chemical reactions and electron transitions.
This is because the radiative field changes much more slowly than hydrogen transitions occur, in general.

\subsection{Bound-Free Transition} 
\label{THI}
Transition between the bound and free states also has a great effect on the radiation field, 
which includes ionization of hydrogen and recombination of free electrons with hydrogen. 
The temporal change in the spectral energy density due to the bound-free transition is given by
\begin{equation}
\label{eq:drhoion}
\nabla \mu I^\mathrm{(u,d)}_{+i}
=  h\nu  \left[\frac{1}{4\pi}  T_{+ i}^\mathrm{A} 
- \frac{1}{c}I_{+ i}^\mathrm{(u,d)} \alpha_{\nu, i} H_i
\right].
\end{equation}
The first term in the bracket on the right-hand side of Eq.~(\ref{eq:drhoion}) represents spontaneous recombination of free electrons, which causes a change in $H_i$ at a rate $T_{+i}^\mathrm{A}$ given by
\begin{equation}
T_{+ i}^\mathrm{A}  = \frac{dA_{+ i}^\mathrm{spn}
 }{dx}\frac{dx}{d\nu} n_\e H_+.
\end{equation}
$A_{+i}^\mathrm{spn}$ is the probability of spontaneous recombination of a free electron with an hydrogen atom by which the electron settles at the energy level $i$.
The express of $A_{+i}^\mathrm{spn}$ is given by 
\cite{Seaton1959} as
\begin{equation}
 \label{eq:Aion}
  \frac{dA_{+ i}^\mathrm{spn}}{dx}
  =
  5.197 \times 10^{-20}
  \epsilon_{i+}^{\frac{3}{2}} 
  e^{\epsilon_{i+}(1-x)} g_i (x) x^{-1} 
  \,\mathrm{m^3 \, s^{-1}}
\end{equation}
where $x \equiv \nu / \nu_i$ ($\nu_i$: the minimum frequency of the $i$th recombination continuum),
$\epsilon_{i+} \equiv x h \nu_i  (k_\mathrm{B}T)^{-1}$
and $g_i (x)$ is the Gaunt factor, for which 
we use the approximated polynomial presented in \citet{Johnson1972}.
We assume that $A_{+i}^\mathrm{spn} = 0$ for $x<1$, because all the electrons are bound by the hydrogen nuclei.
The number densities of hydrogen ions and electrons change with time, which are expressed as
\begin{eqnarray}
\frac{dH_+}{dt} &=& - \frac{dn_\e}{dt}  \nonumber \\
&=&
\sum^{\mathfrak{N}}_{i=1} 
\left[
T^\mathrm{C}_{ i+} - 
\frac{c}{v h\nu}
\int^{\infty}_{\nu_{i}}{ \frac{\ds du_{\e i}}{dt}d\nu}
\right],
\label{eq:dion}
\end{eqnarray}
where
\begin{eqnarray}
\label{eq:icol}
T^\mathrm{C}_{i+}= K_{i+} H_i - K_{+ i}n_\e H_{+}.
\end{eqnarray}
Note that the second term in Eq.~(\ref{eq:dion}) contains both the radiative absorption and stimulated radiation.

The second term in the bracket on the right-hand side of Eq.~(\ref{eq:drhoion}) represents photo-absorptive bound-free transition. 
Its cross-section is given by \cite{Shu1991} as
\begin{equation}
\label{eq:pics}
\alpha_{\nu, i} = i a_1 y_i^{-3} g_i \left(x_i \right),
\end{equation}
where $y_i \equiv h \nu / I_i$ ($I_i$: the ionization energy of the $i$th level hydrogen), and
$a_1$ is a numerical constant (= $7.91\times10^{-22} \mathrm{m^2}$).
In the bound-free transition, we neglect the Doppler broadening (or Doppler shift), because the typical Doppler broadening ratio ($\sim v_\mathrm{th}/c$) is much smaller than the typical continuum width ratio ($\sim \mu v_\mathrm{th}^2/2n_\mathrm{A}h\nu_i$, $n_\mathrm{A}$: the Avogadro constant).
Note that we also consider free electrons that come from atoms other than hydrogen.

\section{Results}
\label{R}
\begin{figure*}[htbp]
	\plottwo{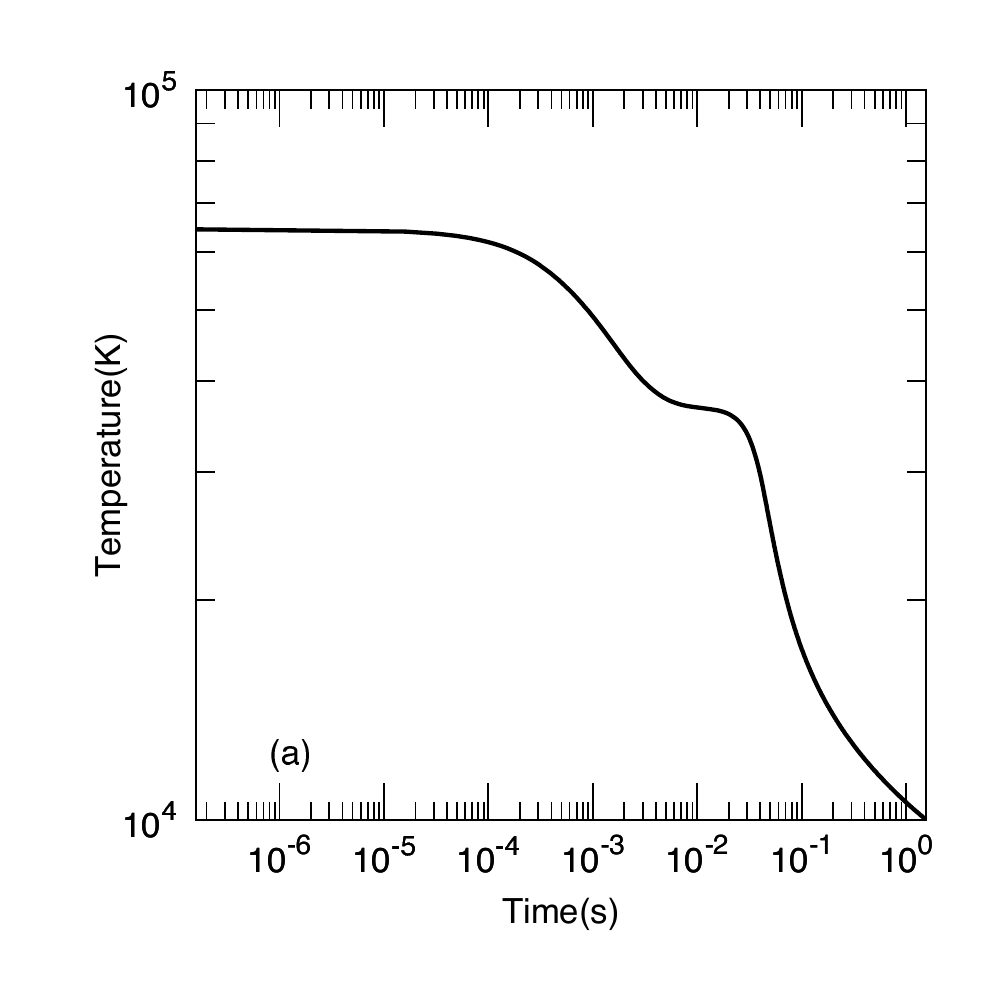}{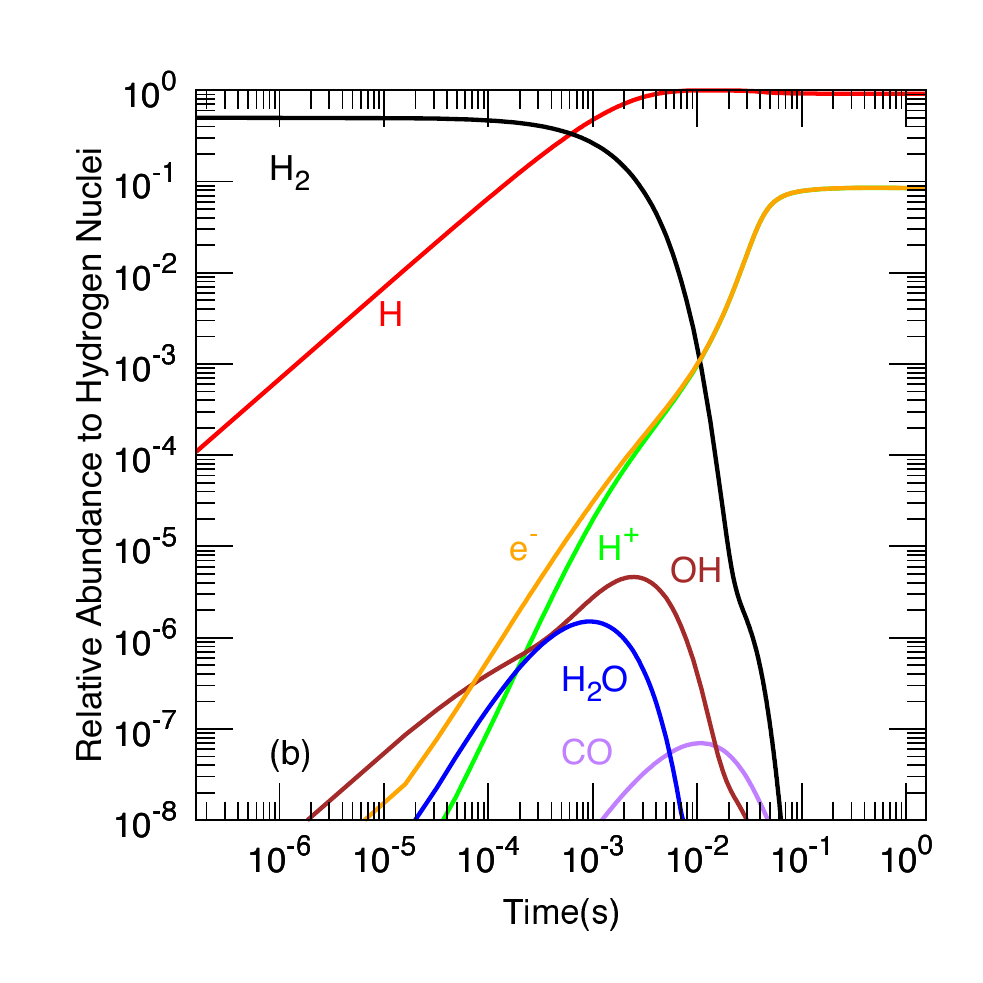}
	\plottwo{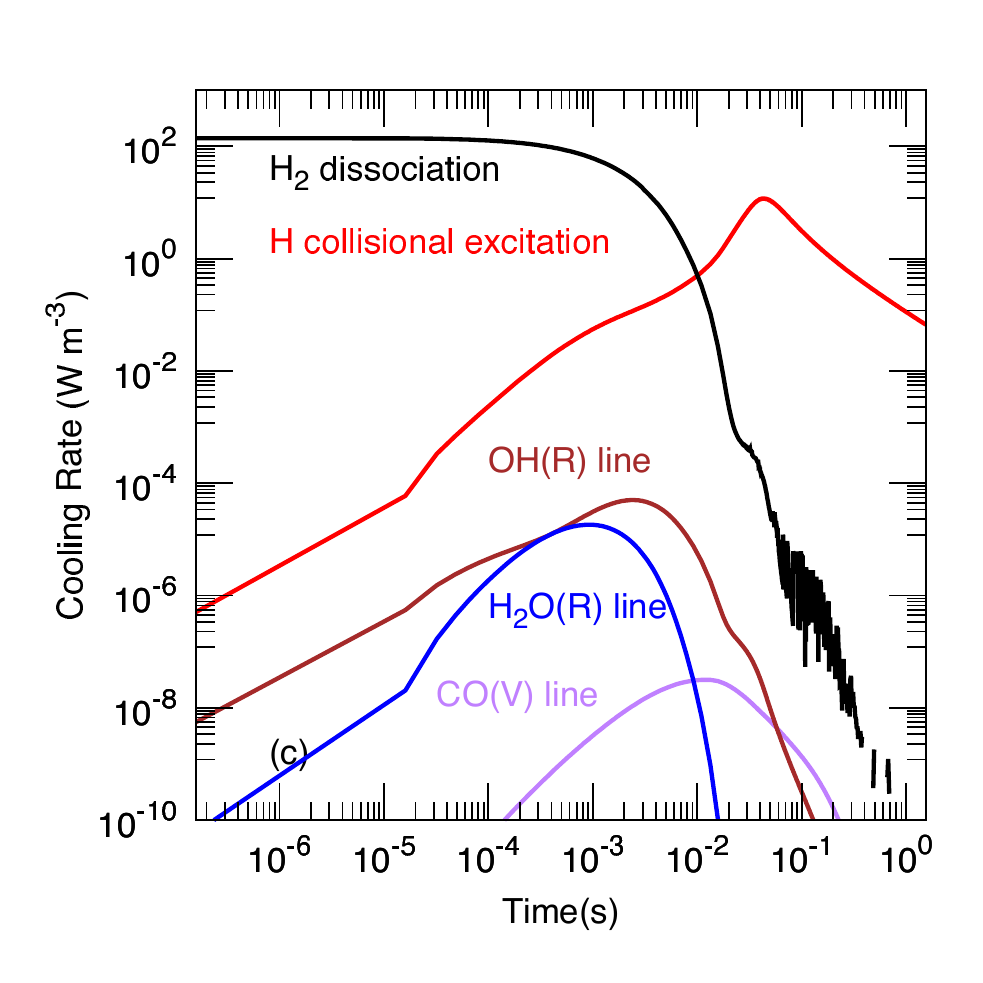}{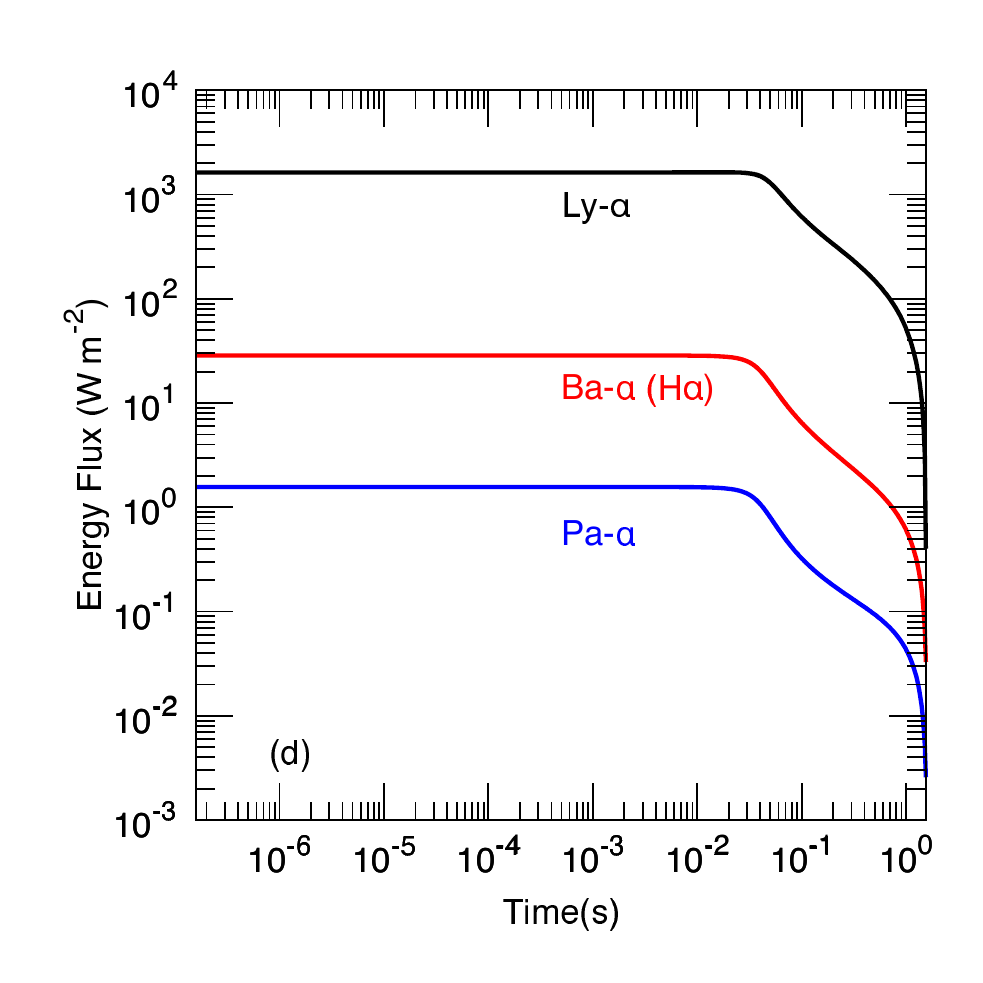}
	\caption{
	Thermo-chemical and radiative processes after the shock front for the preshock velocity $v_0 = 40~\mathrm{km~s^{-1}}$ and the total number density of atomic hydrogen (i.e., protons) per volume $n_{\H,0}=1\times 10^{17}\mathrm{m^{-3}}$. 
	 The panels show temporal changes in (a) the gas temperature; 
	 (b) the numbers of $\H_2$ (black), H (red), $\H^+$ (green), $\e^{-}$ (orange), OH (brown), $\mathrm{H_2O}$ (blue), and CO (purple) relative to $n_{\H,0}$;
	(c) the cooling rate due to $\H_2$ dissociation (black), H collisional excitation (red), OH rotational line cooling (brown), $\mathrm{H_2O}$ rotational line cooling (blue), and CO vibrational radiation (purple);	
	 (d) the energy fluxes of the hydrogen Lyman-$\alpha$ (black), Balmer-$\alpha$ or H$\alpha$(red), and Paschen-$\alpha$ (blue) emissions.
	}
	\label{fig:40_17}
\end{figure*}

Here we present numerical results of the thermo-chemical and radiative processes that the flow undergoes after passing through the shock. 
The input parameters in this flow model include the preshock velocity $v_0$ and the total number density of atomic hydrogen (including all the hydrogen nuclei in the molecules such as $\mathrm{H}_2$ and $\mathrm{H_2 O}$) $n_\mathrm{H,0}$.
As the fiducial case, we adopt $v_0=40~\mathrm{km~s^{-1}}$ and $n_\mathrm{H,0}=1\times 10^{17} \mathrm{m^{-3}}$, for which we investigate in section \ref{FC}.
Then, we show results for denser gas ($n_\mathrm{H,0}=1\times 10^{20} \mathrm{m^{-3}}$) in section~\ref{denser} and for higher velocity ($v_0=90~\mathrm{km~s^{-1}}$) in section~\ref{higher}, followed by a parameter study in section~\ref{PS}.

\subsection{Fiducial Case}
\label{FC}
Figure~\ref{fig:40_17} shows temporal changes in postshock quantities for $v_0=40~\mathrm{km~s^{-1}}$ and $n_{\H,0}=1 \times 10^{17}$~$\mathrm{m}^{-3}$ \footnote{These values are not always the typical ones for accreting gas giants. We have chosen them for validating our numerical model by comparing it with the \citet{Iida+2001} model.}. 
Since the change of the individual fluid parcel is observed, the horizontal axis also corresponds to the spatial coordinate $z$ (i.e., the Lagrangian coordinate), which means that Fig.~\ref{fig:40_17} shows the vertical distribution of the quantities below the shock front.

First, the gas temperature changes as follows (see panel~[a]): 
At $t$ = 0, the temperature reaches as high as $6.9\times 10^4$~K, because of shock heating. 
The flowing gas remains at that temperature until $t \simeq$ $2\times 10^{-4}$~s.  
Then, the gas cools down to $4 \times 10^4$~K in about $1\times 10^{-2}$~s 
and remains at that temperature until $t\simeq$ $2\times 10^{-2}$~s. 
After that, cooling occurs again. 
This change in temperature is related to chemical reactions, as follows. 
As shown in panel~(b), H continues to form by dissociation of $\H_2$ in the first $3\times10^{-3}$~s.
Concurrently, H is being ionized to $\H^+$ and $\e^-$. 
The decrease of $H_2$ is linked to the increase of H and e$^-$,
since the dissociation of $\mathrm{H}_2$ is due mainly to collision with H or e$^-$. 
In panel~(c), it turns out that the gas cools by two different dominant processes: 
The first cooling phase ($t \lesssim10^{-2}$~s) is governed by $\H_2$ dissociation,
whereas the second phase ($t \gtrsim 10^{-2}$~s) is controlled by collisional excitation of H. 
Molecular line emission has little contribution to cooling at high temperatures shown in Fig.~\ref{fig:40_17}, 
because molecules such as CO and OH are present only in small amounts. 

\begin{figure}[htbp]
	\plotone{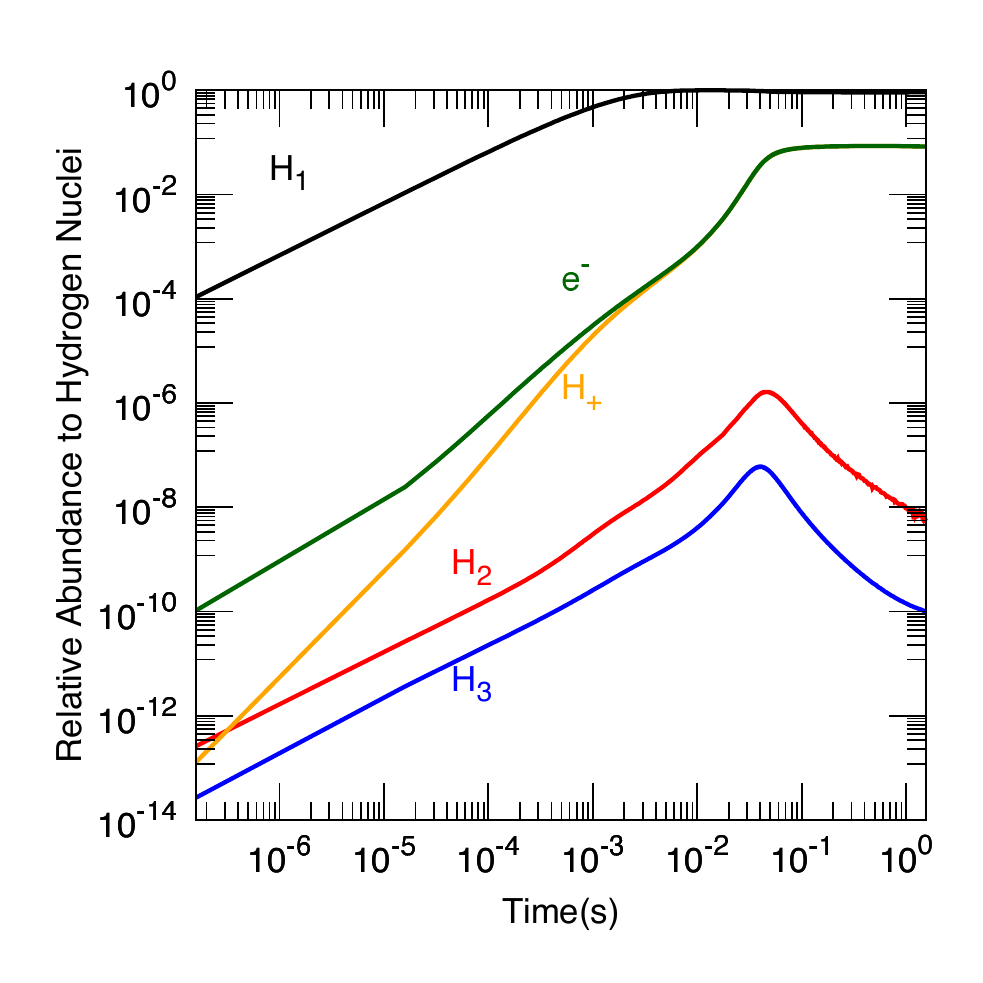}
	\caption{
	The electron level population after the shock front for $v_0=40~\mathrm{km~s^{-1}}$ and $n_{\H,0}=1 \times 10^{17}\mathrm{m^{-3}}$: The relative number of isolated hydrogen atoms whose principal quantum number is 1 (black), 2 (red), and 3 (blue), ionized H (orange), and free electron (green).
	}
	\label{fig:H40_17}
\end{figure}

Panel~(d) shows the upward energy fluxes of the hydrogen Lyman-$\alpha$ (black), Balmer-$\alpha$ (red), and Paschen-$\alpha$ (blue) emission.
All the fluxes increase monotonically upstream (from right to left in panel [d]). 
In this case, the line emission occurs predominantly at $t\simeq2\times 10^{-2}$~s.
This is because the gas is relatively cool in the deep regions ($t\gtrsim 2\times 10^{-2}$~s), while the number of electrons, which excite hydrogen, is too small in the shallow regions ($t\lesssim2\times 10^{-2}$~s).
This can be understood also from Fig.~\ref{fig:H40_17} that shows the temporal change in the number of isolated hydrogen atoms (relative to the total number of hydrogen nuclei) with principal quantum number $i_q$ of 1 (black), 2 (red), and 3 (blue) and also the number of hydrogen ions (orange) and electrons (green).
The numbers of hydrogen ions and electrons increase until $t \simeq$ 2-3 $\times$~$10^{-2}$~s and then become nearly constant. 
As seen in Fig.~\ref{fig:40_17}a, the temperature immediately after shock is high enough to dissociate and ionize hydrogen, producing free electrons. 
Those electrons collide with and excite hydrogen atoms. 
Thus, as electrons increase, excited hydrogen atoms increase.
The hydrogen excitation, on the other hand, results in cooling the gas, which then leads to reducing the number of free electrons. 
However, because hydrogen ion recombination proceeds only slowly ($>10$~s), the abundance of $\H^+$ and $e^-$ is almost constant for $t \gtrsim 2\times 10^{-2}$~s in Fig. \ref{fig:H40_17}.
For $t\gtrsim 3\times 10^{-2}$~s, since gas temperature drops, the number of hydrogen atoms of $i_q \geq 2$ naturally decreases.

\begin{figure*}[htbp]
	\plottwo{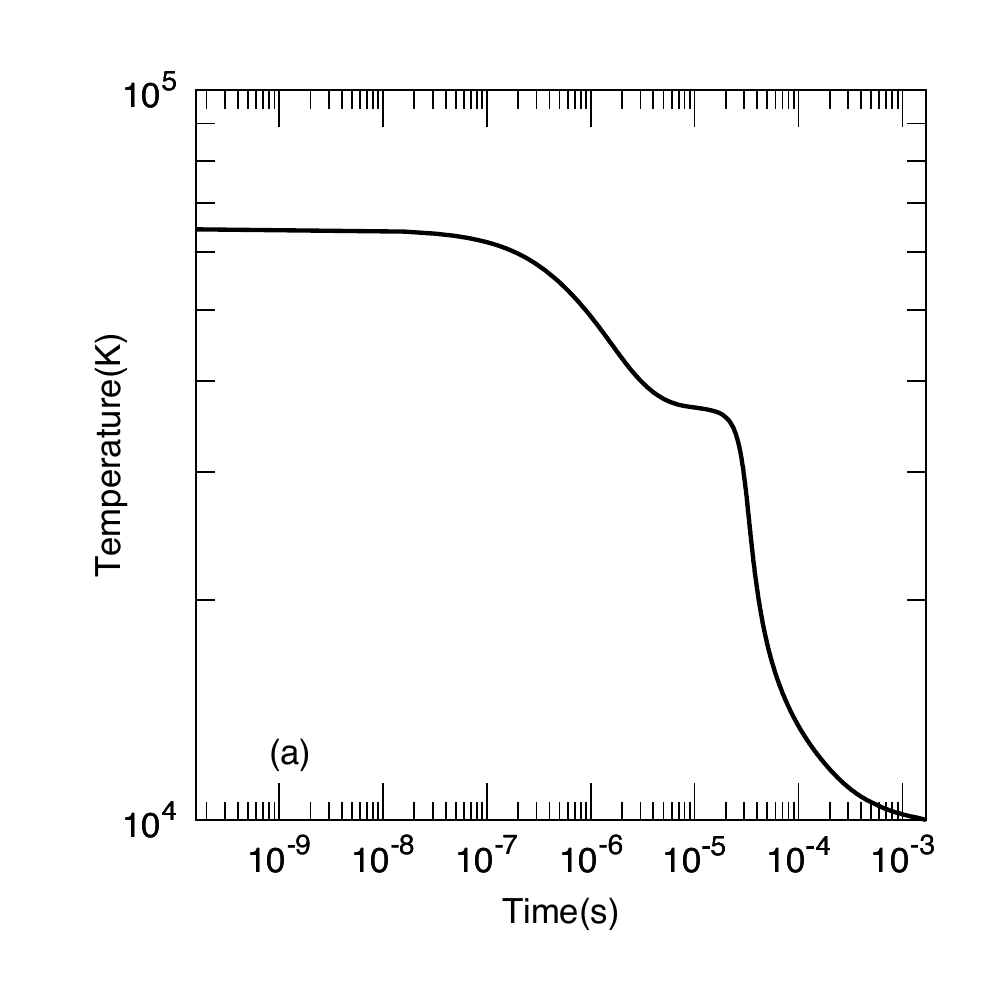}{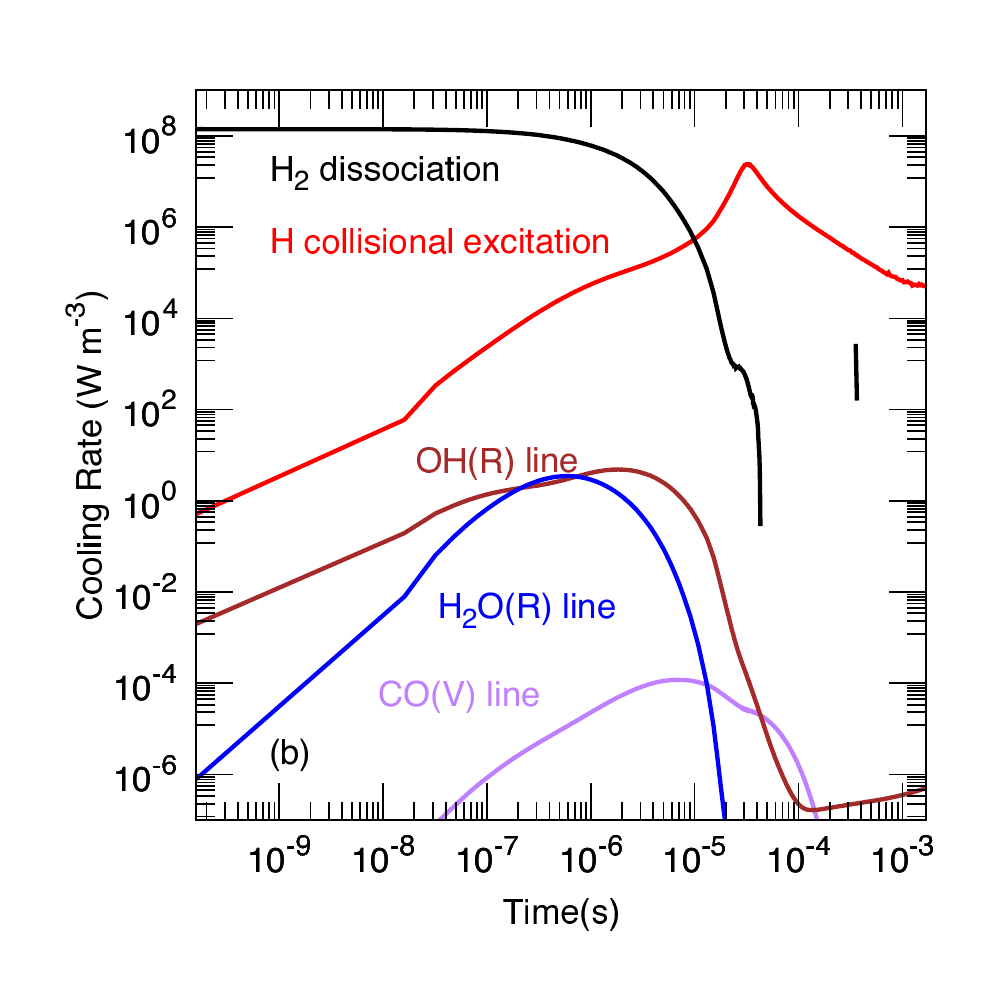}
	\plottwo{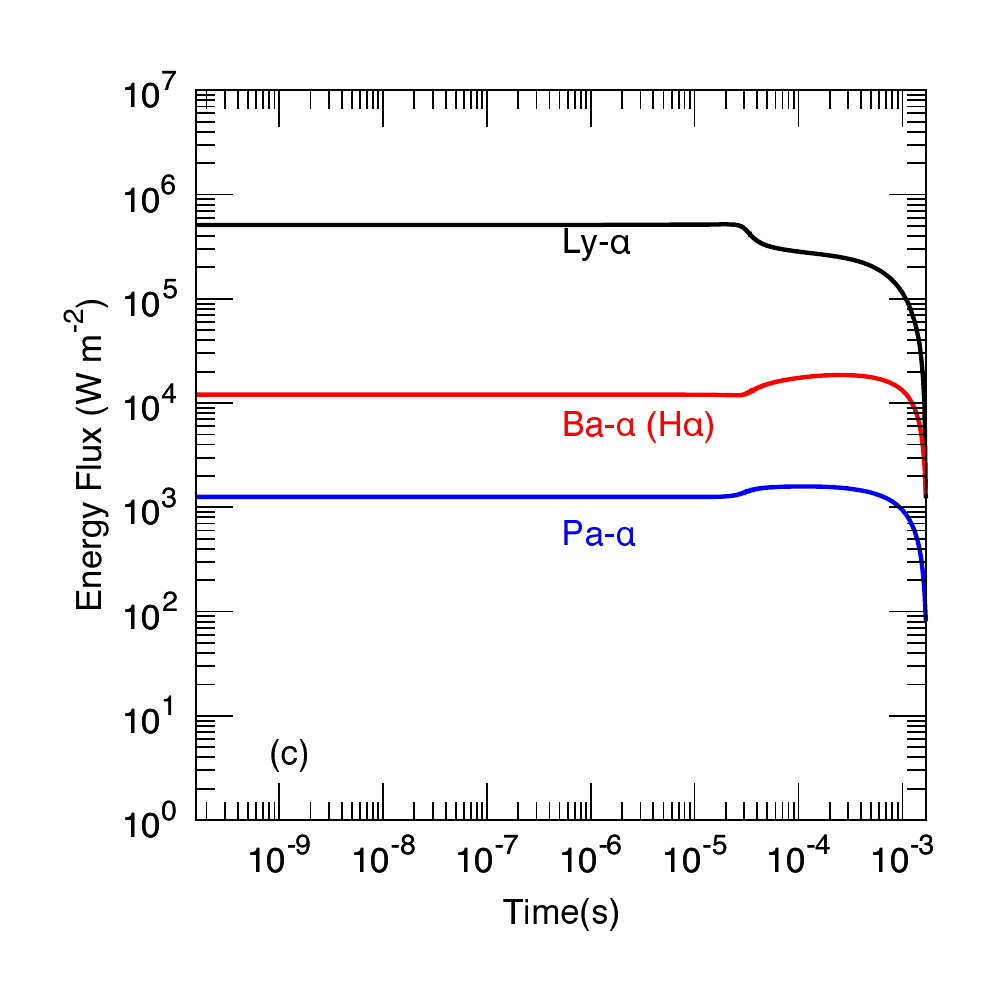}{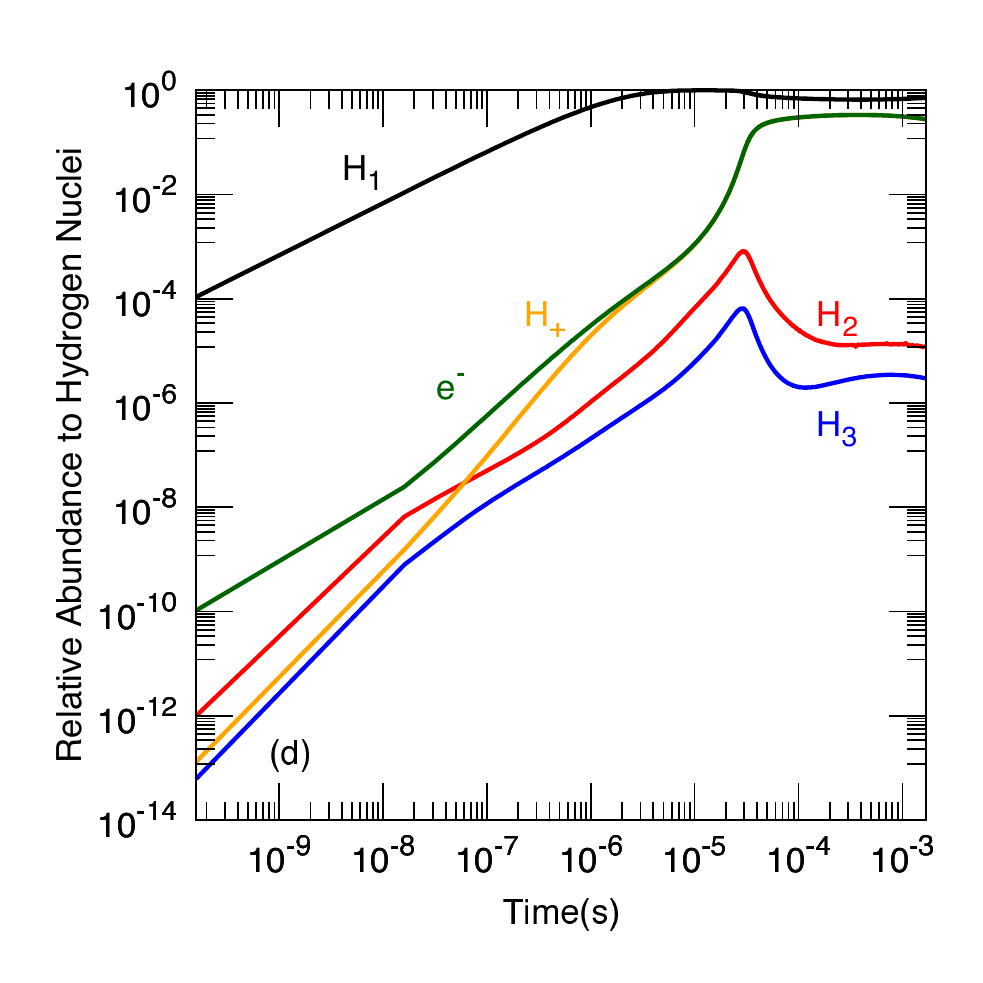}
	\caption{	
	Thermal and radiative processes after the shock front for $v_0 = 40~\mathrm{km~s^{-1}}$ and $n_{\H,0}$ = $1 \times 10^{20}\mathrm{m^{-3}}$.
	The panels show temporal changes in
	(a) the gas temperature;
	(b) the cooling rate due to $\H_2$ dissociation (black), H collisional excitation (red), OH rotational radiation (brown), $\mathrm{H_2O}$ rotational radiation (blue), and CO vibrational radiation (purple);	
	(c) the upward radiative energy flux of hydrogen Lyman-$\mathrm{\alpha}$ (black), Balmer-$\alpha$ or H$\alpha$ (red), and Paschen-$\alpha$ (blue);
	(d) the number of isolated hydrogen atoms whose principal quantum numbers are 1 (black), 2 (red), and 3 (blue) and the number of ionized H (orange). 
	}
	\label{fig:H40_20}
\end{figure*}

\subsection{High Density Case}
\label{denser}
Figure~\ref{fig:H40_20}a, b, c, and d are the same as Fig.~\ref{fig:40_17}a, c, d, and Fig.~\ref{fig:H40_17}, respectively, but for $n_{\H,0}=1\times10^{20} \mathrm{m}^{-3}$.
In a denser gas, because of frequent collisions, collisional excitation and de-excitation of hydrogen take place more frequently, so that the postshock processes driven by collisions (e.g., temperature drop) proceed on shorter timescales.
On the other hand, the timescale of spontaneous de-excitation is independent of gas number density.
Thus, hydrogen is more excited and ionized in a denser gas.
A larger number of electrons also lead to further excitation of hydrogen.

In Fig.~\ref{fig:H40_20}c, the profile for the Lyman-$\alpha$ shows a different feature from those for the other two lines.
At $t\sim 10^{-3}$~s, the Lyman-$\alpha$ flux is on the order of $10^5 ~\mathrm{W~m^{-2}}$,
which is high enough that the absorption and emission of Lyman-$\alpha$ balance with each other there.
(Note that energy density per wavelength, instead of energy flux, is high enough, exactly to say.)
This means that the gas is optically thick with respect to the Lyman-$\alpha$ radiation.
Thus, for $t\lesssim 10^{-3}$~s, the emission rate of the Lyman-$\alpha$ line radiation is determined locally by the abundance of the first-excited hydrogen  $H_2$ (red line in Fig.~\ref{fig:H40_20}d), which is the source of Lyman-$\alpha$ photons, and the ground-state hydrogen, $H_1$ (black line in Fig.~\ref{fig:H40_20}d).
Indeed, the Lyman-$\alpha$ flux increases sharply with decreasing time at $t\sim3\times 10^{-5}$~s, which corresponds to the peak time for $H_2$, around which $H_1$ also increases moderately.
The reason why the Lyman-$\alpha$ \textit{increases} with decreasing time is that $H_2$ changes \textit{more rapidly} than $H_1$.

No similar feature is seen for the Balmer-$\alpha$ and Paschen-$\alpha$ lines in Fig.~\ref{fig:H40_20}c.
This is because those fluxes are too low for the gas to be optically thick.
Note that the Balmer-$\alpha$ flux slightly decreases with decreasing time around $t\sim 3 \times 10^{-5}$~s, which corresponds to the peak time for $H_3$ (see Fig.~\ref{fig:H40_20}d).
Unlike in the case of Lyman-$\alpha$, $H_2$ decreases more gently than $H_3$ with decreasing time.
This is because absorption of Lyman-$\alpha$, which suppresses decrease in $H_2$, is greater than that of Balmer-$\alpha$ or Lyman-$\beta$, which suppresses decrease in $H_3$.

As shown in Fig.~\ref{fig:H40_20}d, the change in $H_2$ (red line) shows a somewhat different feature from that in the fiducial case (red line in Fig.~\ref{fig:H40_17}).
From $3\times 10^{-5}~\mathrm{s}$ to $\sim10^{-4}~\mathrm{s}$, temperature drops and thus the number of the excited hydrogen atoms decreases, same as in the fiducial case.
However, after that (i.e., $t \gtrsim10^{-4}~\mathrm{s}$), the decrease in the number of excited hydrogen seems to be rather moderate.
This is because excitation due to absorption of the line radiation from downstream compensates for collisional de-excitation.
In the fiducial case, namely optically thin case, the absorptive excitation is much less efficient than the collisional excitation.
The $H_3$ (blue line) shows the similar feature with $H_2$ (red line), but increases a bit around $t\sim10^{-4}$~s.
In this case, the number of the second-excited hydrogen $H_3$ is supported by both of the de-excitation from upper levels, especially the ionised state, and line absorptive excitation.
Therefore, $H_3$ increases slightly around $t\sim10^{-4}$~s, though the collisional de-excitation dominates over the collisional excitation in that temperature range.

\begin{figure*}[htbp]
        \plottwo{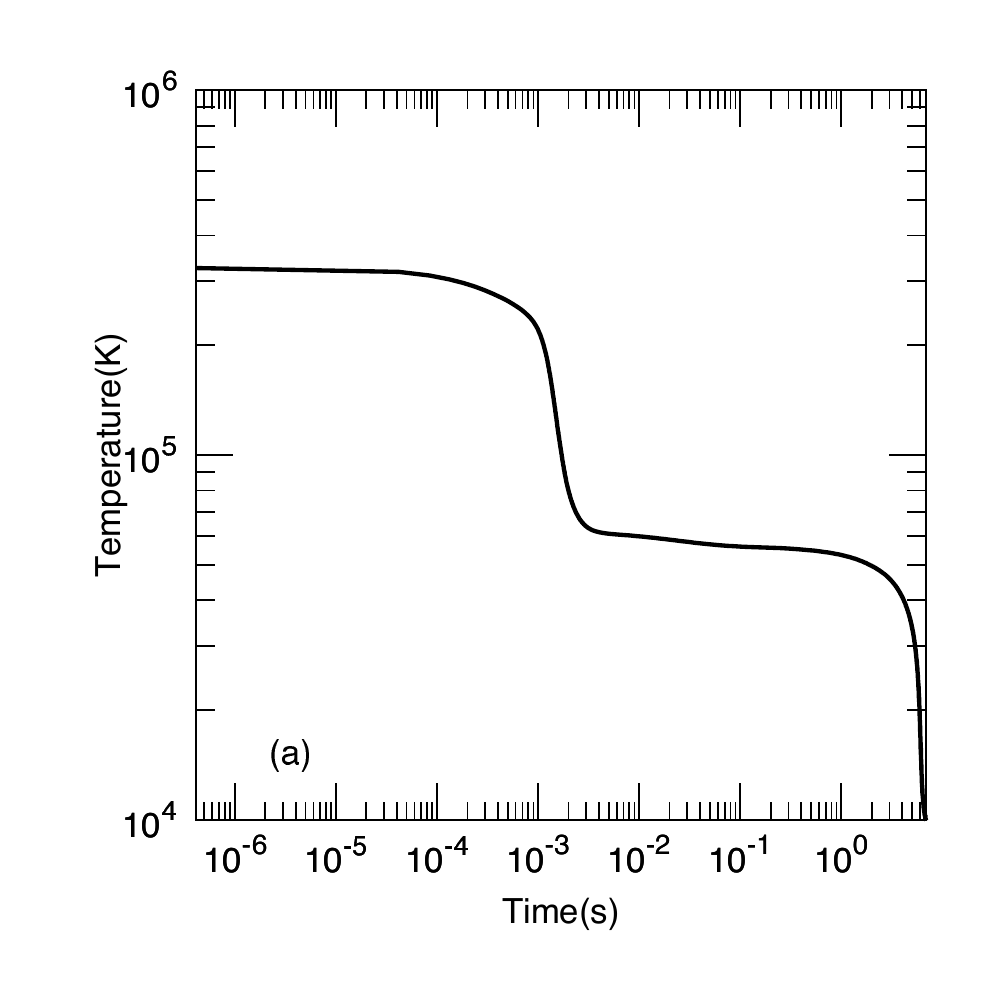}{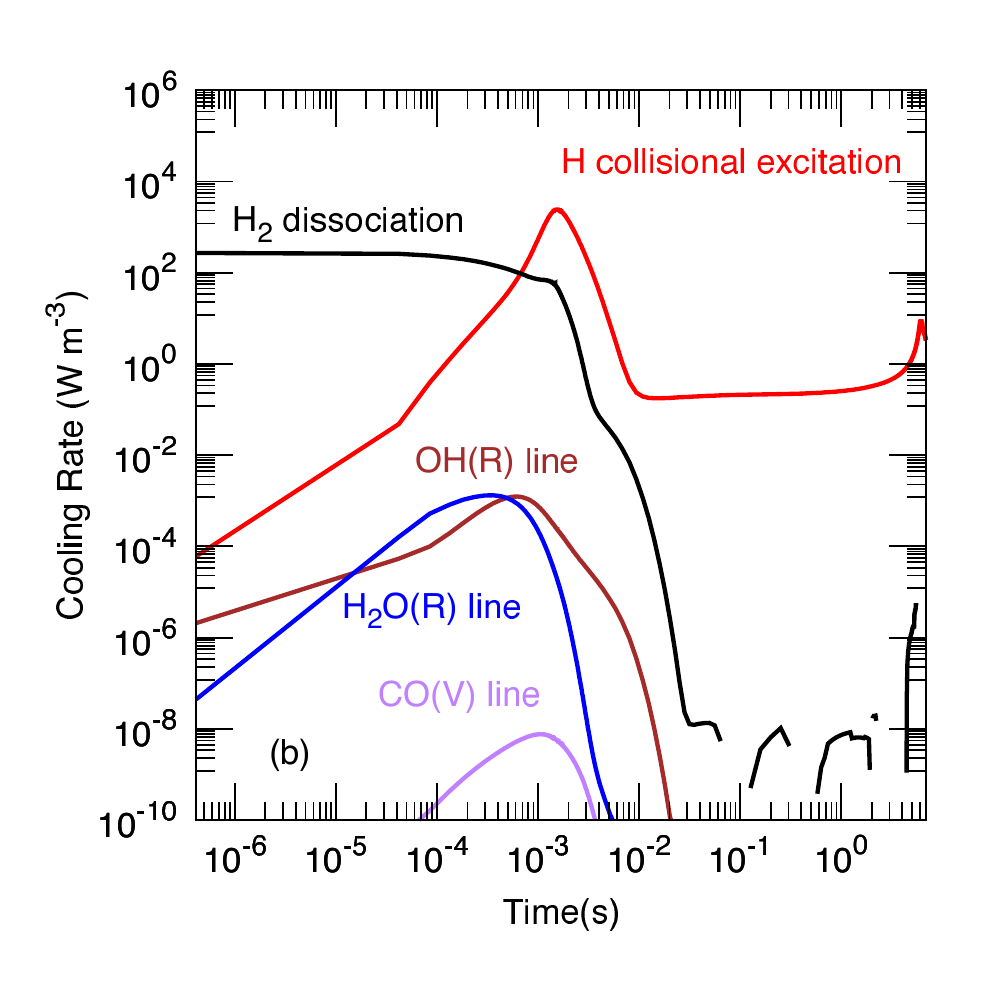}
        \plottwo{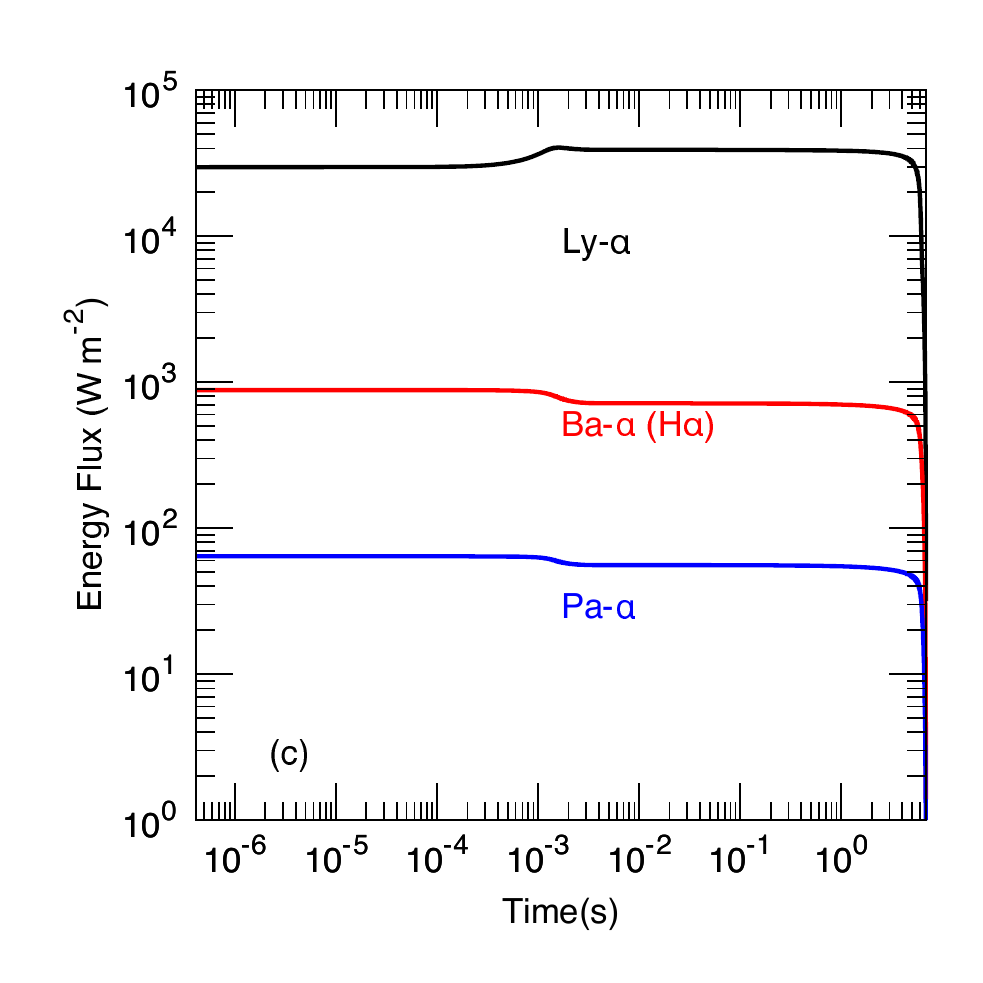}{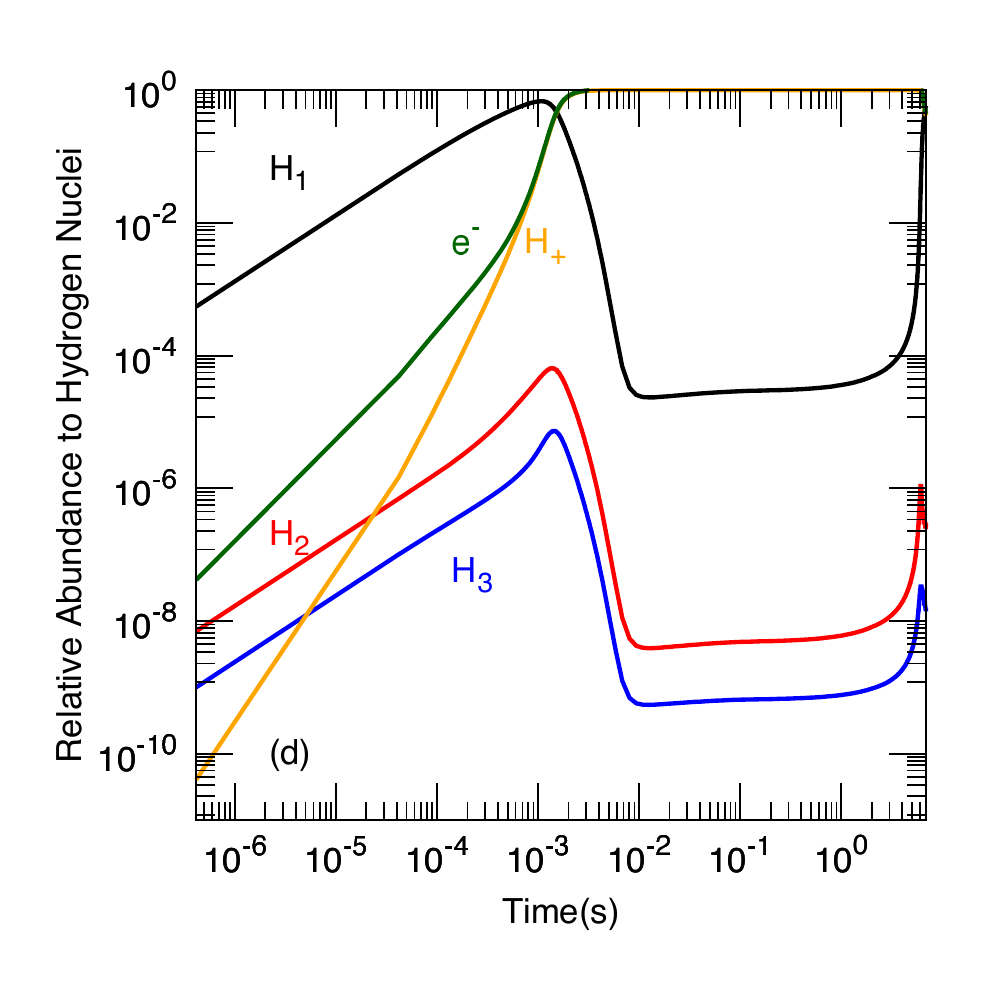}
	\caption{
	Same as Fig.~\ref{fig:H40_20}, but for $v_0=90~\mathrm{km~s^{-1}}$ and $n_{\H,0}=1\times10^{17}~\mathrm{m^{-3}}$.
	}
	\label{fig:H90}
\end{figure*}

\subsection{Higher Velocity Case}
\label{higher}

Figure~\ref{fig:H90} is the same as Fig. \ref{fig:H40_20} but for a higher preshock velocity $v_0=90~\mathrm{km~s^{-1}}$.
In this case, the gas temperature exceeds $1 \times 10^5$~K immediately after shock (see panel [a]). 
Because of such high temperature, hydrogen is ionized almost completely (see panel [d]), and thus the number of hydrogen nuclei with electrons (i.e., neutral hydrogen) is much smaller than in the case of lower $v_0$. 
Because of almost no neutral hydrogen, namely no strong coolant, the cooling timescale is quite long in the highly ionized region ($t\gtrsim 2\times 10^{-3}$~s).
Once the gas temperature goes below a certain value at $\sim 5$~s, the ionization rate drastically drops and neutral hydrogen is reproduced (see panel [d]). 
Thus, hydrogen line radiation is generated in that region (see panel [c]).
The rare neutral hydrogen region means optically thin for hydrogen lines.
That is why hydrogen line energy flux mainly changes before ($t \lesssim10^{-2}$~s) and after ($t \gtrsim1$~s) the high ionization region.

\subsection{Parameter Study}
\label{PS}

\begin{figure*}[htbp]
	\plottwo{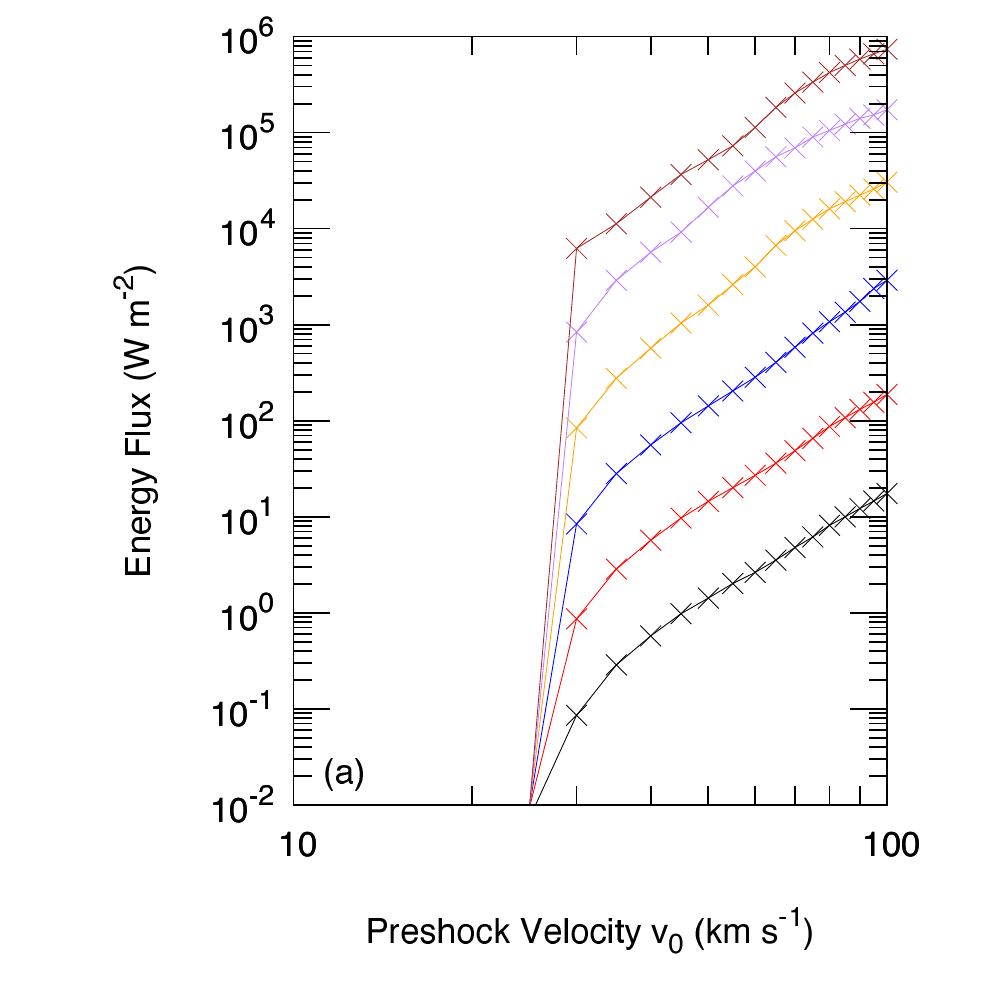}{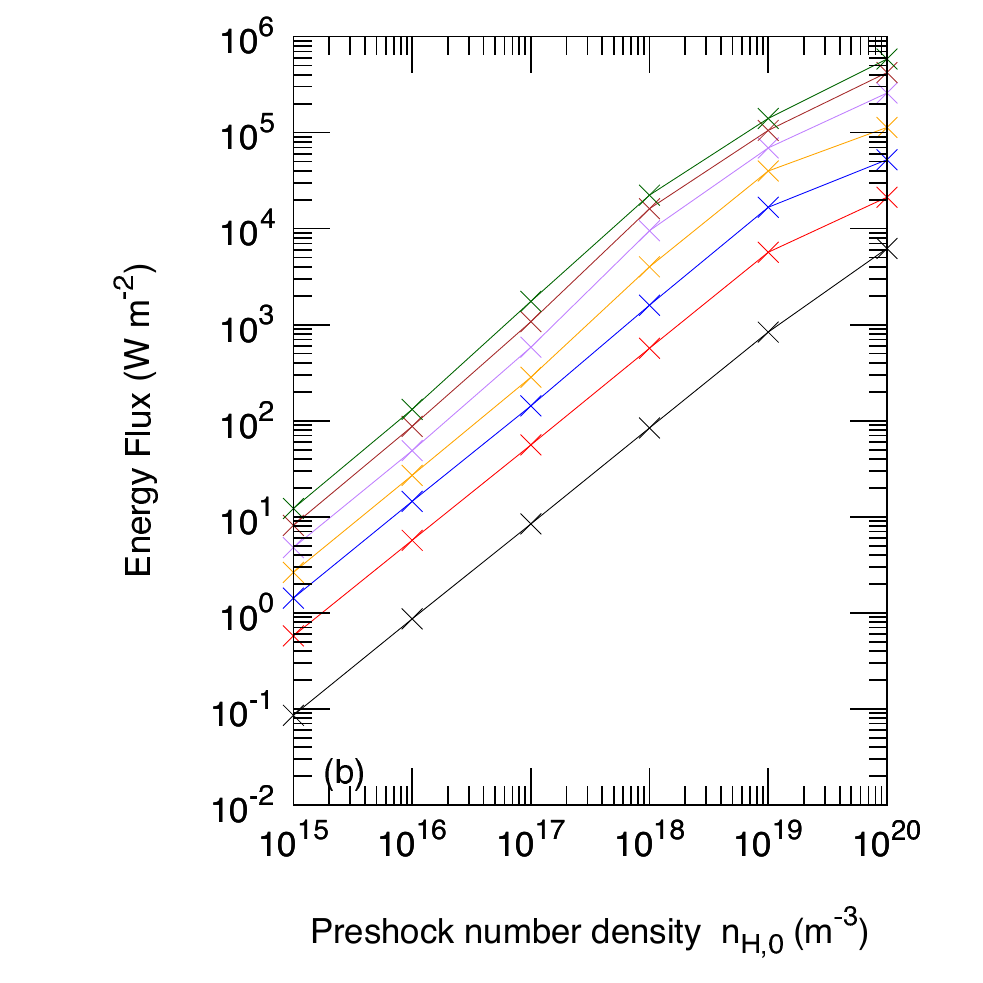}
	\plottwo{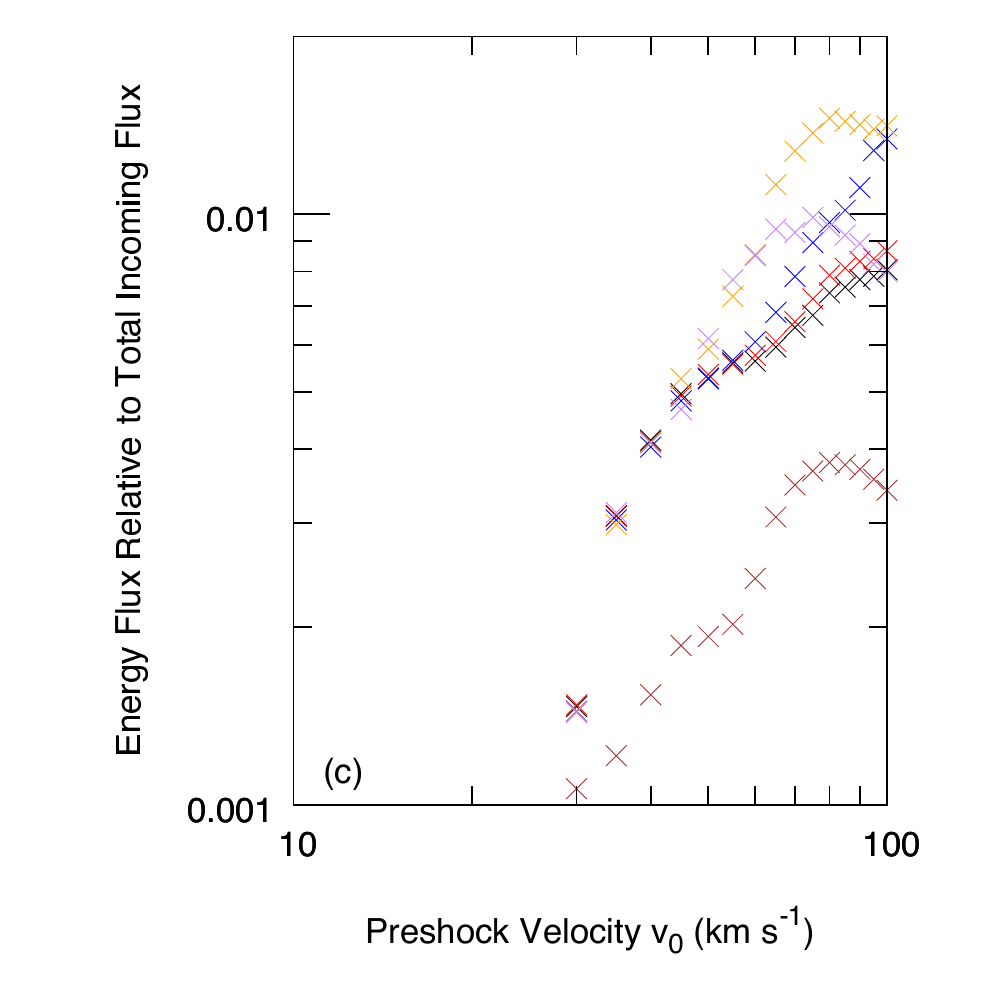}{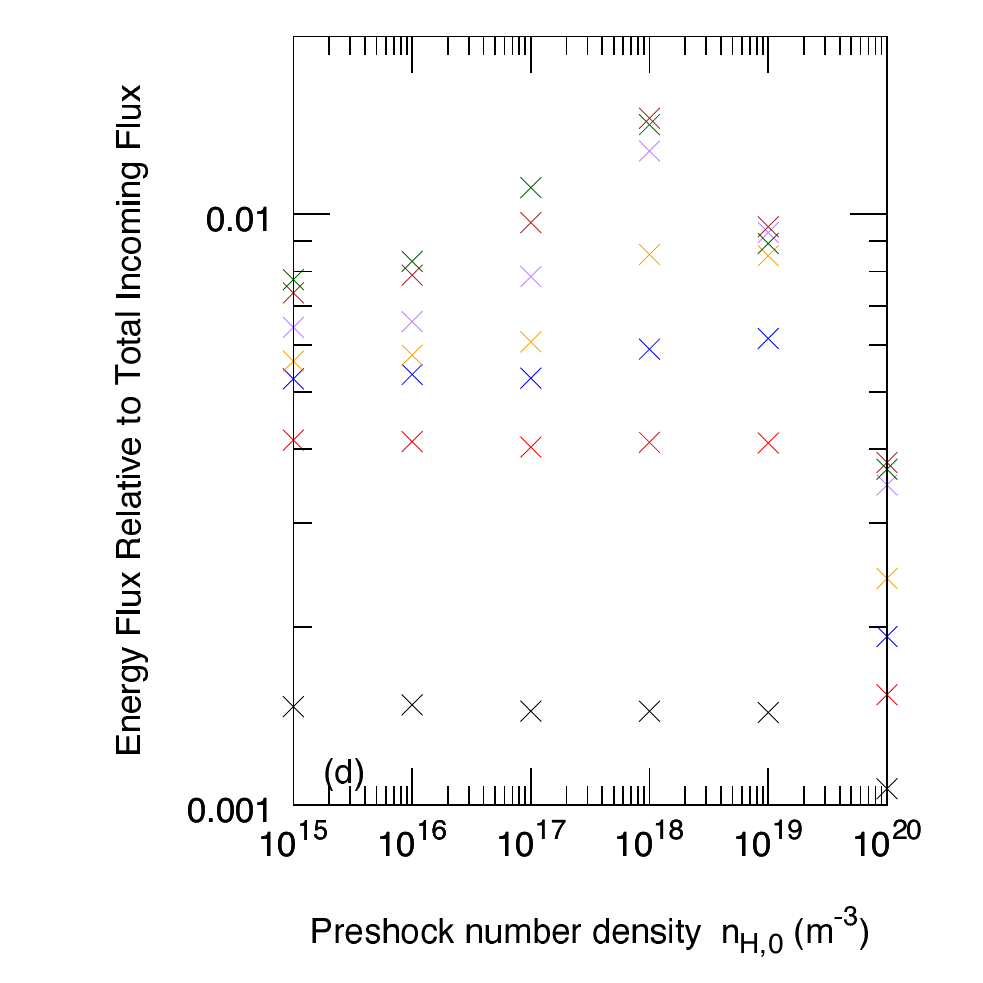}
	\caption{	
The energy flux of the Balmer-$\alpha$ line (H$\alpha$) radiation at the shock front (i.e. the surface of the circum-planetary disk) against the preshock velocity $v_0$ (panels (a) and (c)) and the total number density of atomic hydrogen (including all the hydrogen nuclei) $n_\mathrm{H,0}$ (panels (b) and (d)). 
The upper two panels are the same as the lower two, respectively, but the latter shows the energy flux relative to the total incoming energy flux at the shock front (i.e., $y_\mathrm{t}\mu v_0^3 n_\mathrm{H,0}/2$ where $y_\mathrm{t}n_\mathrm{H,0}$ is the total number of particles and $\mu$ is the mean molecular weight).
In the left two panels, each color shows each choice of $n_\mathrm{H,0}$; $10^{15}$~m$^{-3}$ (black), $10^{16}$ m$^{-3}$ (red), $10^{17}$ m$^{-3}$ (blue), $10^{18}$ m$^{-3}$ (orange), $10^{19}$ m$^{-3}$ (purple), and $10^{20}$ m$^{-3}$ (brown).
Also in the right two panels, each color shows each choice of $v_0$; $30~\mathrm{km~s^{-1}}$ (black), $40~\mathrm{km~s^{-1}}$ (red), $50~\mathrm{km~s^{-1}}$ (blue), $60~\mathrm{km~s^{-1}}$ (orange), $70~\mathrm{km~s^{-1}}$ (purple), $80~\mathrm{km~s^{-1}}$ (brown), and $90~\mathrm{km~s^{-1}}$ (green).
}	
	\label{fig:ps}
\end{figure*}

In Fig.~\ref{fig:ps}, we show the dependence of the Balmer-$\alpha$ line energy flux on (a) the preshock velocity $v_0$ and  (b) the total number density of hydrogen nuclei $n_\mathrm{H,0}$, respectively. 
In Appendix, we also show the dependences regarding other lines such as Balmer-$\beta$, Paschen-$\alpha$, and Paschen-$\beta$\footnote{Those profiles would be similar even at infinity except for Lyman-$\alpha$, because hydrogen is in the form of H$_2$ and excited H rarely exists above the shock front. As for Lyman-$\alpha$, absorption in the interstellar medium modifies the profiles.}.
As seen in panel~(a), for $v_0 < 30$~km~s$^{-1}$, the Balmer-$\alpha$ line flux is quite low.
This is because almost all the energy of shock heating is consumed for dissociation of hydrogen molecules.
For $v_0 \geq 30$~km~s$^{-1}$, the energy flux is found to be nearly proportional to $v_0^4$ in panel (a) and to $n_{\H,0}$ in panel (b). 
Intuitively, however, the energy flux is proportional to $v_0^3 n_\mathrm{H,0}$, because the kinetic energy that the flowing gas has before shock is $v_0^3 n_\mathrm{H,0}/2$. 
This holds true for Lyman-$\alpha$, but for other lines, we have to take into account the effect of absorption of radiation propagated from downstream, as described below.

\begin{figure*}
\plottwo{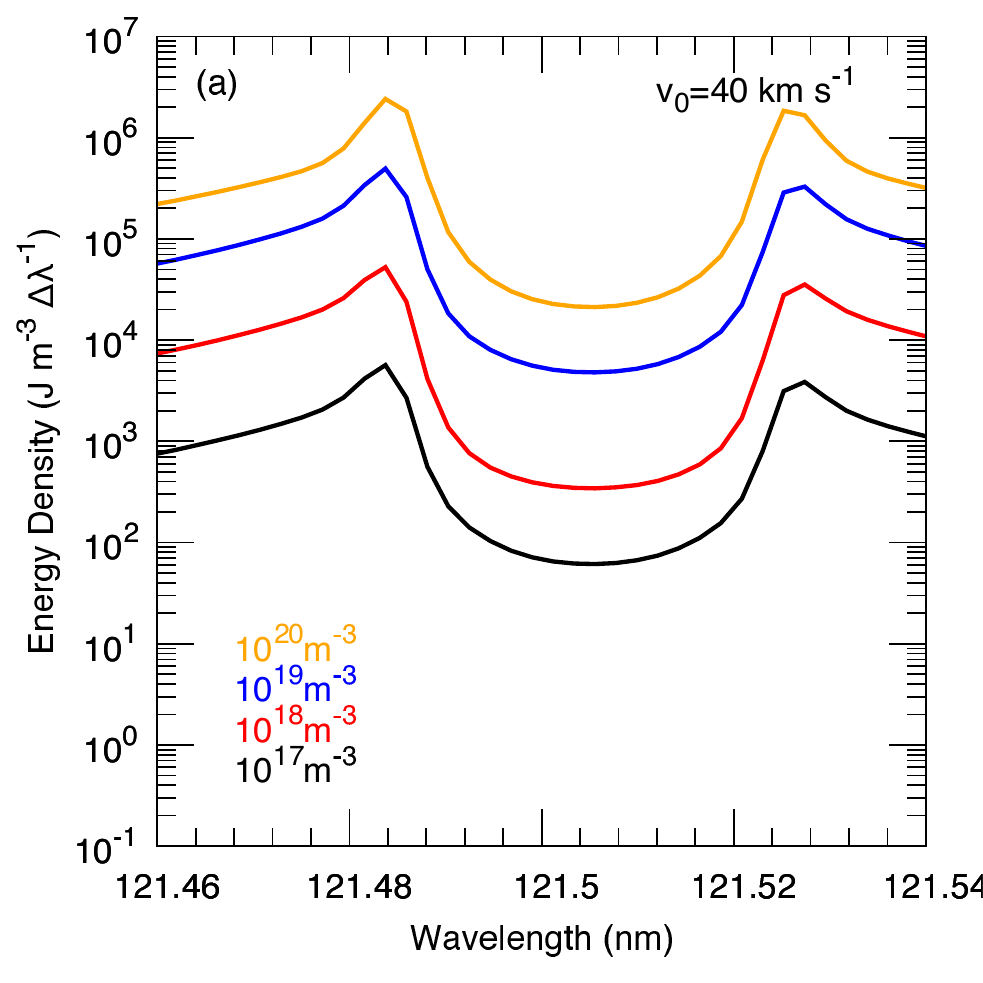}{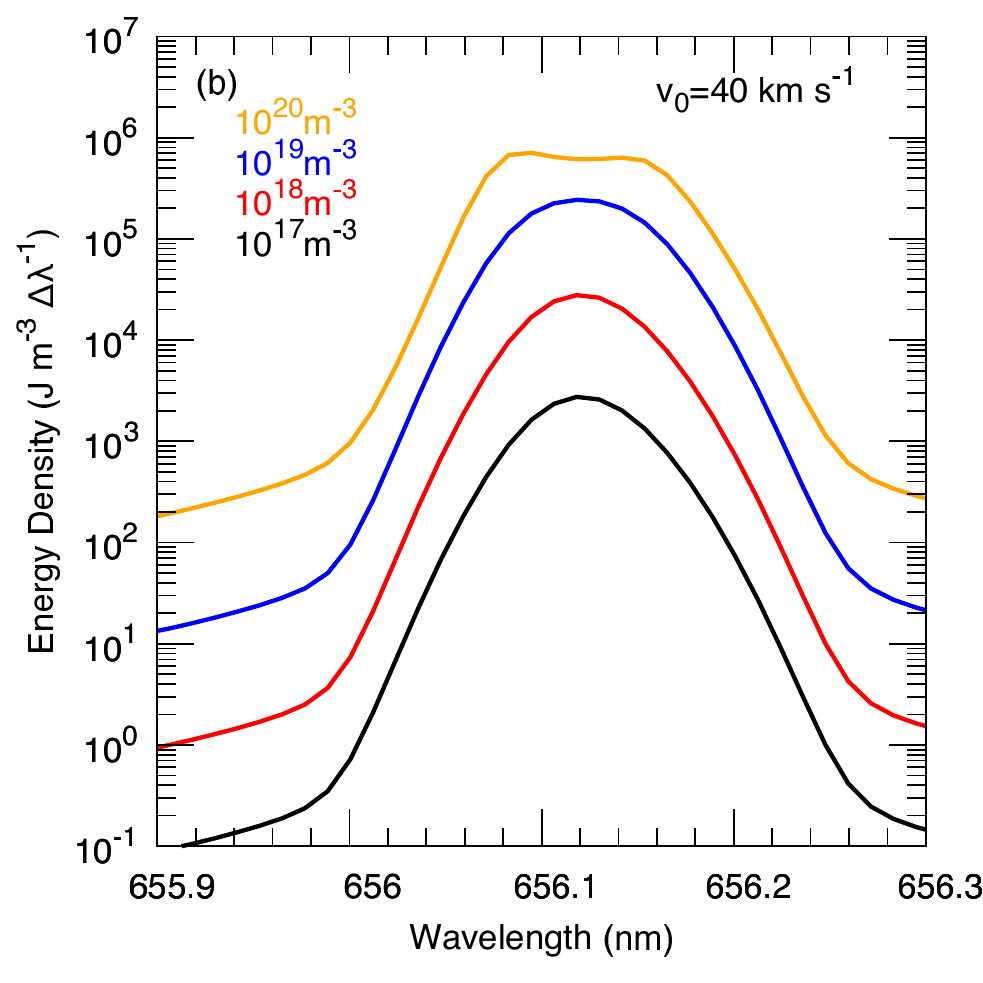}
\plottwo{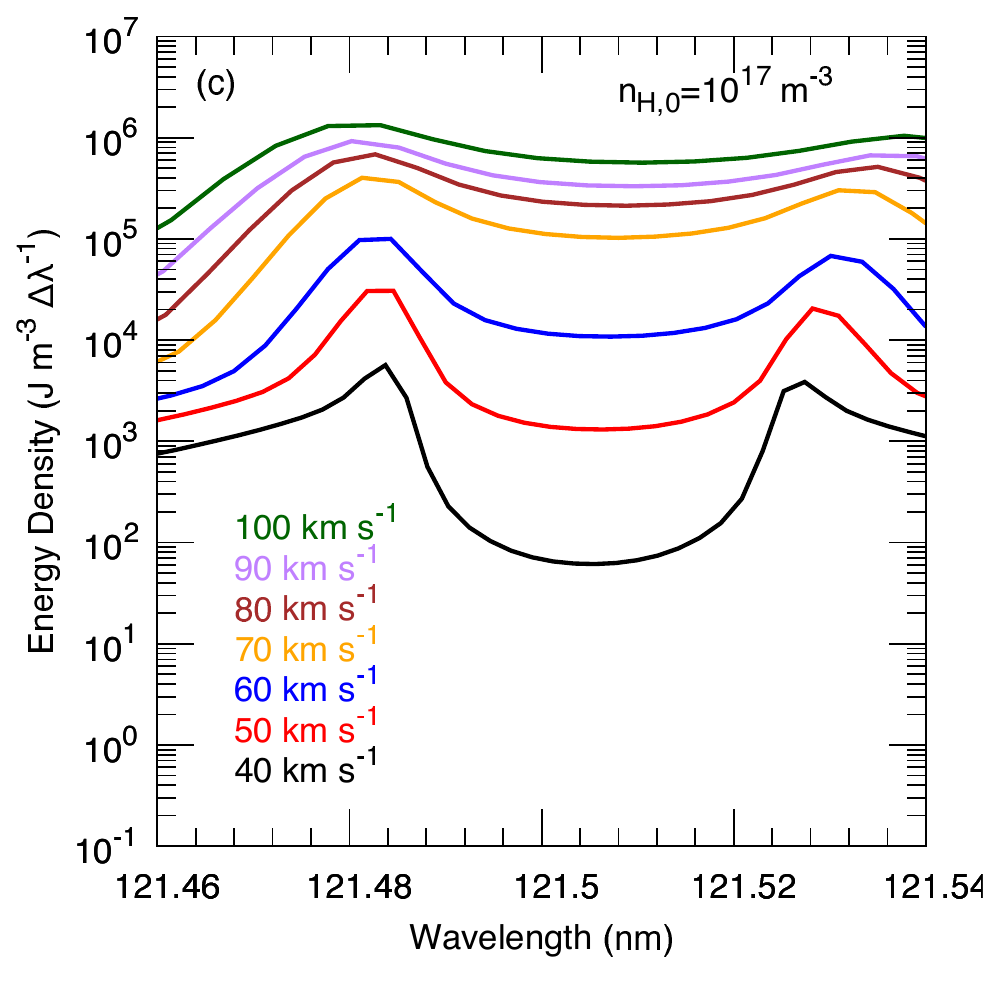}{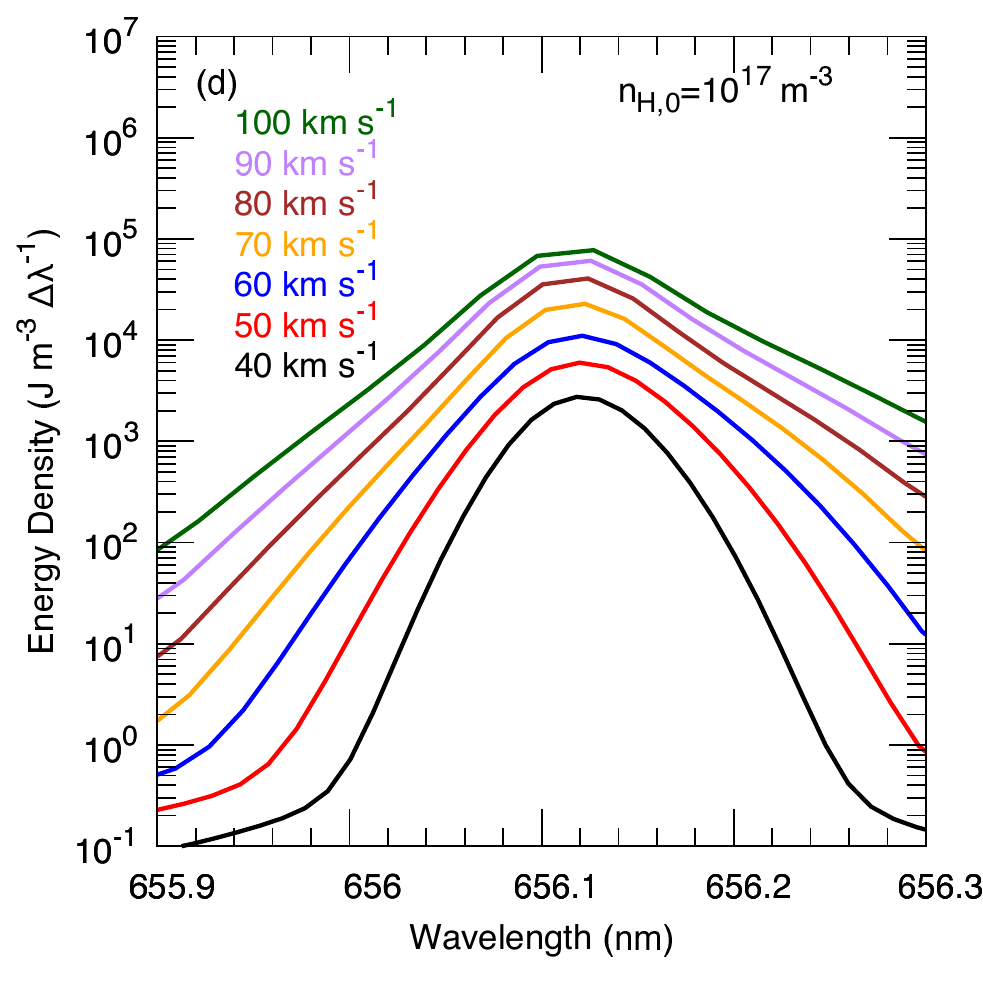}
\caption{
Spectral energy density profile of hydrogen Lyman-$\alpha$ line (left panels) and Balmer-$\alpha$ line (right panels) emitted upward at the shock front.
In panels (a) and (b), the preshock velocity $v_0 = 40$~km~s$^{-1}$.
The lines are colored according to the total number density of hydrogen nuclei; $n_\mathrm{H,0}$ = $10^{17}$ m$^{-3}$ (black), 10$^{18}$ m$^{-3}$ (red), 10$^{19}$ m$^{-3}$ (blue), and 10$^{20}$ m$^{-3}$ (orange).
Panels (c) and (d) are for the case of $n_{\H,0}=10^{17} \mathrm{m}^{-3}$ and $v_0 = 40$~km~s$^{-1}$ (black), 
50~km~s$^{-1}$ (red), 
60~km~s$^{-1}$ (blue), 
70~km~s$^{-1}$ (orange), 
80~km~s$^{-1}$ (brown), 
90~km~s$^{-1}$ (purple), and 
100~km~s$^{-1}$ (green).
}
\label{fig:profile}
\end{figure*}

Figure~\ref{fig:profile} shows the profiles of the spectral energy density of the Lyman-$\alpha$ and Balmer-$\alpha$ lines at the shock front.
In the upper row panels, those for different choices of $n_{\H,0}$ are presented for $v_0=40$~km~s$^{-1}$, while in the lower panels, those for different choices of $v_0$ are presented for $n_{\H,0}=10^{17}~\mathrm{m^{-3}}$.
The Gauss profile caused by the Doppler broadening is seen near the line center, while the Lorenz profile mainly caused by the natural broadening is seen far from the line center.
The borders between them are at $\sim 121.48$~nm and $\sim121.53$~nm for Lyman-$\alpha$ (in panel [a]) and $\sim656.00$~nm and $\sim 656.26$~nm for Balmer-$\alpha$ (in panel [b]), respectively.
In panel (b), for $n_\mathrm{H,0}= 10^{17}$ to $10^{19} \mathrm{m^{-3}}$, the energy density is found to be nearly proportional to $n_\mathrm{H,0}$ as a whole.

Basically, the number densities of all the species are proportional to $n_\mathrm{H,0}$. 
However, 
because high density leads to high cooling rate and their relationship is almost linear, hydrogen line emission occurs in a shallower region, the depth of which is inversely proportional to $n_\mathrm{H,0}$. 
Consequently, the column density of the emission region hardly depends on $n_\mathrm{H,0}$. 
Thus, another reason is needed for explaining the linear dependence of Balmer-$\alpha$ emission on $n_\mathrm{H,0}$.
The electron level distribution is roughly in equilibrium between the radiative de-excitation and collisional excitation. 
Since the former and latter are proportional to $n_\mathrm{H,0}$ and $n_\mathrm{H,0}^2$, respectively, 
the number ratio of the emitter to absorber of hydrogen lines and thus the emitted line flux are nearly proportional to $n_\mathrm{H,0}$. 
On the other hand, comparing the profiles for $n_\mathrm{H,0}=10^{19}$ and $10^{20}~\mathrm{m^{-3}}$ in panel (b), one realizes that the energy density is proportional to $n_\mathrm{H,0}$ far from the center, but not near the center.
This comes from photo-absorption (see also Fig.~\ref{fig:H40_20}c).
Since the gas flow velocity is higher where photons are absorbed than where photons are emitted, the center of absorption feature is shifted to longer wavelength relative to the center of emission feature.
This is why the left peak is higher than the right peak of the yellow line in panel (b).
For Lyman-$\alpha$ in panel (a), all the profiles show the same feature more obviously.
Since the Lyman-$\alpha$ is optically thick enough that the energy density is determined locally, the energy density near the line center is proportional to $H_2/H_1$ ratio in the Lyman-$\alpha$ photosphere (or optical depth $\tau=1$ plane).

As shown in panels (c) and (d) of Fig.~\ref{fig:profile}, 
as $v_0$ increases, the line width becomes larger, because of increase in temperature,
and the energy flux becomes larger, because of increase of excited hydrogen.
For the Lyman-$\alpha$ in Fig.~\ref{fig:profile}c, the absorption feature becomes weaker with increasing $v_0$ in the case for $n_0=10^{17}~\mathrm{m^{-3}}$.
This is because the gas temperature and the $H_2/H_1$ ratio, which determines the energy density locally in optically thick case, come to be higher with increase of $v_0$ at the $\tau=1$ surface of Lyman-$\alpha$.

\section{Discussion}
\label{D}
\subsection{Hydrogen Line Luminosity from Accreting Planets}

As described in Introduction, 
\citet{Sallum+2015} reported on the detection of a source of H$\alpha$ (or Balmer-$\alpha$) emission in the circum-stellar disk of LkCa15.
In order to put a physical interpretation on the origin and intensity of this emission and constrain the ranges of the planet mass and disk gas density, we integrate the line emission flux obtained above throughout the CPD surface.
The total emergent flux of hydrogen line emission from the CPD is given by
 \begin{eqnarray}
 \label{eq:luminosity_r}
 L&=&\int_{r_\mathrm{P}}^{\infty} 2F(v_0(r),n_{\H,0}) 2\pi r dr,
\end{eqnarray} 
where $F(v_0,n_{\H,0})$ is the emergent intensity per unit area from the shock front obtained in our simulation,
$r$ is the radial distance from the center of the protoplanet, and 
$r_\mathrm{P}$ is the protoplanet radius, which is assumed to be twice the Jupiter's radius.
Here we have assumed an axisymmetric CPD.

\begin{figure}[htbp]
\includegraphics[scale=0.6, bb=9 9 425 425]{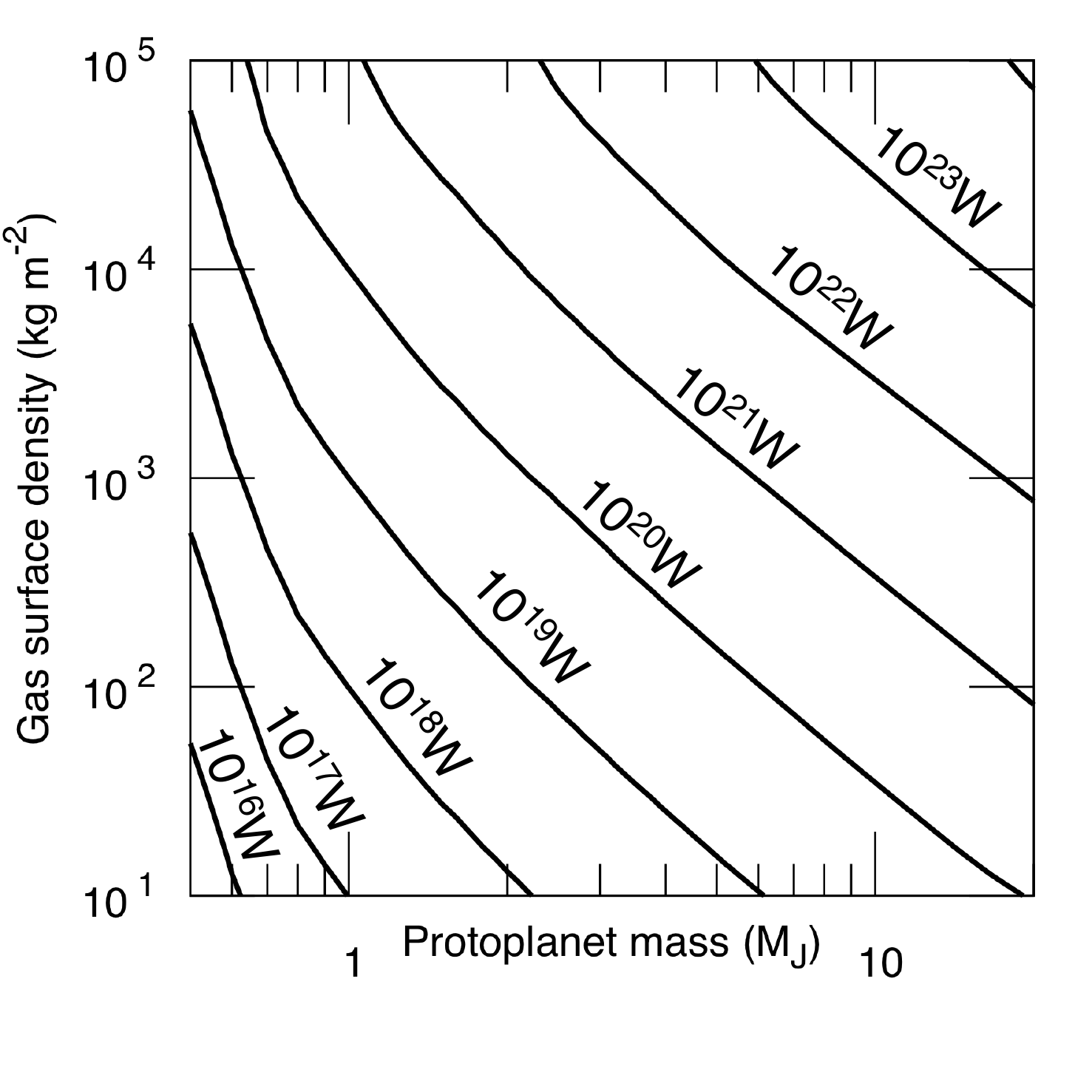}
\caption{ 
Contour plot of the H$\alpha$ luminosity (see Eq.~[\ref{eq:luminosity_r}] and [\ref{eq:nonv}] for the definition) versus the protoplanet mass in Jupiter mass $M_\mathrm{J}$ and the surface density of protoplanetary disk gas at $14.7$AU around a $1 M_\odot$ protostar.
}
\label{fig:con_Ha}
\end{figure}

We obtain the functions of $v_0 (r)$ and $n_{\rm H, 0} (r)$ from 3D hydrodynamic simulations by \citet{Tanigawa+2012}, as follows.
The accreting gas flows vertically onto the CPD surface.
Since the gravity from the CPD is much weaker than that from the protoplanet, we assume that $v_0$ is the free fall velocity to the protoplanet,
\begin{eqnarray}
v_0&=&\sqrt{\frac{2GM_\mathrm{p}}{r}},
\end{eqnarray}
where $G$ is the gravitational constant and $M_\mathrm{p}$ is the protoplanet mass.
The number density $n_\mathrm{H, 0}$ is derived from the condition of steady flow, namely, the mass flux $J$ is equal to 
$\mu y_\mathrm{t} n_\mathrm{H, 0} v_0$, where $y_\mathrm{t}$ is the particle number density normalized by $n_\mathrm{H,0}$. 
According to the 3D simulations \citep[see Fig.~13 of][]{Tanigawa+2012}, 
$J \sim 5 \Sigma_0 \Omega_\mathrm{K}$ in the inner CPD, 
where $\Sigma_0$ is the unperturbed surface density of CSD and $\Omega_\mathrm{K}$ is the Keplerian angular velocity around the central star.
Thus, $n_{\rm H, 0}$ is given from the relation $\mu y_\mathrm{t}n_{\rm H, 0} v_0 = 5 \Sigma_0 \Omega_\mathrm{K}$ as
 \begin{equation}
 \label{eq:nonv}
 n_\mathrm{H,0}=\frac{5\Sigma_0} {\mu y_\mathrm{t}} \sqrt{ \frac{M_* r}{2M_\mathrm{p}a^3} },
\end{equation}
where $M_*$ is the mass of the central star and $a$ is the orbital semi-major axis of the forming gas giant.

Using $M_*=1M_\odot$ and $a=14.7$~AU (and $\mu = 2.4 \times 10^{-27}$~kg) from \citet{Sallum+2015}, we integrate Eq.~(\ref{eq:luminosity_r}) 
for various values of $v_0$ and $n_{\rm H, 0}$ and get the H$\alpha$ luminosity contour shown in Fig.~\ref{fig:con_Ha}.
\citet{Sallum+2015} estimated the H$\alpha$ luminosity of LkCa15b to be $2.3\times 10^{22}$~W. 
Note that they took interstellar extinction into account in deriving this luminosity:
To be exact, one also has to consider extinction that occurs in the vicinity of the planet from this value.
This would be, however, small:
The gas falling on CPD hardly scatters H$\alpha$ photons because it contains few first-excited hydrogen atoms.
The disk gas surface density was estimated by \citet{Marel+2015} to be $165~\mathrm{kg~m^{-2}}$ at 14.7~AU from LkCa15.
Applying those two values to Fig.~\ref{fig:con_Ha}, we find that the mass of LkCa15b is more than 20~$M_\mathrm{J}$, which is out of the planet mass range.
This is inconsistent with the mass of LkCa15b, 10~$M_\mathrm{J}$, inferred from the Ks-band observation \citep{Sallum+2015}. 

This contradiction argues for the need for further observations of this object and detailed theoretical investigation of the accretion process of massive gas giants. 
Recent IR observation by \citet{Thalmann+2016} reported on the detection of scattered radiation from the outer disk around LkCa15, which might imply that the H$\alpha$ detected by \citet{Sallum+2015} was also the scattered one.
On the other hand, regarding the gas accretion model, the numerical factor of 5 used in Eq.~(\ref{eq:nonv}) is valid when the planet's Hill radius is equal to the disk scale height \citep{Tanigawa+2012}. 
This factor may depend on planet mass.
According to \citet{Tanigawa+2002}, the gas accretion rate is proportional to $M_\mathrm{P}^{1.3}$.
If we assume that the mass flux onto the circum-planetary disk is proportional to this gas accretion rate and apply such a relation to Fig.~\ref{fig:con_Ha}, the mass of LkCa15b is estimated at $12M_\mathrm{J}$ for $\Sigma_0=165~\mathrm{kg~m^{-2}}$.
However, this is to be examined, because this estimation includes no information of the 3D distribution of gas around the protoplanet.

\citet{Uyama+2017} observed the protostar TW Hya in the Paschen-$\beta$ line with Keck/OSIRIS.
TW Hya is known to have a multi-ring (or multi-gap) protoplanetary disk \citep[e.g.][]{Calvet+2002, Menu+2014},
suggesting the presence of accreting gas giants in the gaps of the disk.
No Paschen-$\beta$ excess was, however, detected in 5$\sigma$ detection limit.
The detection limits correspond to $2.4\times 10^{17}$~W and $6.3\times10^{16}$~W for the two large disk gaps at 25~AU and 95~AU, respectively.
Adopting the gas surface densities of 270~kg m$^{-2}$ and 4.9~kg m$^{-2}$ from the photoevaporating-disk model of \citet{Gorti+2011}, 
we estimate the upper limits of the protoplanet masses to be $\sim$ 2~M$_\mathrm{J}$ and $\sim$ 8~M$_\mathrm{J}$, respectively.
Note that the latter estimate is different from that in \citet{Uyama+2017}, because we have assumed constant mass flux at the surface of the circum-planetary disk, in contrast to \citet{Uyama+2017} who assumed constant gas density.

For a set of mass accretion rate and protoplanet mass, our estimate of hydrogen line emission is weaker by a few orders of magnitude than that from the empirical relationship used in stellar accretion context \citep{Gullbring+1998}.
This is due to the differences in preshock velocity and gas accretion feature.
Because of weak gravity, the preshock velocity is lower and then postshock gas is cooler in planetary accretion than in stellar accretion.
The vertical accretion flow which causes strong hydrogen line emission accounts for only a small fraction of the whole accreting gas, while most of the accreting gas falls onto the outer regions of the CPD, where the flow velocity is too slow for the gas to be hot enough to generate hydrogen line emission. 
Although of great importance are applying our model to stellar accretion and then comparing stellar and planetary accretion, we need more complicated and time-consuming calculations where absorption of hydrogen line radiation by the preshock gas will likely make a great contribution. Such comparison will be done in our future study.

\subsection{Caveats}
\subsubsection{Effects of Magnetic Field}
The deep interior of accreting gas giants is hot enough that hydrogen is ionized and convecting \citep[e.g.][]{Bodenheimer+Pollack1986}. 
This suggests that accreting gas giants have intrinsic magnetic fields.
Provided incoming gas is partially ionized, magnetic waves can affect the strength of shockwaves, which is weakened,
if the propagation velocity of magnetic waves, $v_\mathrm{A}$, (termed the magnetosonic velocity) is larger than the flow velocity, $v_0$.
In terms of the magnetic field $B$, this condition is expressed as
 \begin{eqnarray}
 B  &\gg& 2 \sqrt{\pi \rho}  v_0  \nonumber \\
 & \sim& 2.2
 \left( \frac{x_\mathrm{t}}{1.4}  \right)
 \left( \frac{n_{\H, 0}}{10^{17} \mathrm{m^{-3}}} \right)^{\frac{1}{2}} 
 \left( \frac{v_0}{40~\mathrm{km \, s^{-1}}} \right) 
 \mathrm{T},
\end{eqnarray}
where $x_\mathrm{t}$ is the molar mass (in gram).
Given even the current surface magnetic field of the Sun is less than 1~T, this condition is unlikely to be satisfied in the case of gas giants.

In the post-shock regions where hydrogen line radiation is generated, partial ionization occurs obviously. Thus, the magnetic fields could influence the hydrodynamic and radiative processes there.
From momentum conservation, the dynamical pressure is converted not only into the thermal pressure but also into the magnetic pressure. 
This means that the existence of magnetic field leads to reducing the gas number density for a given temperature.
The characteristic number density $n_\mathrm{m}$ and temperature $T_\mathrm{m}$, respectively, above and below which the magnetic pressure dominates over the thermal one is given by \citep{HM79}:
\begin{eqnarray}
n_\mathrm{m}&=&\sqrt{8\pi \rho}\frac{n_0v_0}{B_\perp} \nonumber \\
&\sim& 3.1 \times 10^{17} \left( \frac{x_\mathrm{t}}{1.4}\right)^{\frac{1}{2}}
\left( \frac{n_0}{10^{17}\mathrm{m^{-3}}}\right)^{\frac{3}{2}}\nonumber \\&&
\left( \frac{v_0}{40~\mathrm{km\,s^{-1}}}\right)
\left( \frac{B_\perp}{\mathrm{1 T}}\right)^{-1}
\mathrm{m^{-3}},
\\
T_\mathrm{m}&=& \frac{\rho_0 v^2_0}{y_\mathrm{t}n_\mathrm{m} k_\mathrm{B}} \nonumber \\
&\sim& 1.5 \times 10^5
\left( \frac{x_\mathrm{t}}{1.4}\right)^\mathrm{-\frac{1}{2}}
 \left( \frac{y_\mathrm{t}}{0.6} \right)^{-1}
 \left( \frac{n_0}{10^{17}\mathrm{m^{-3}}} \right)^{-\frac{1}{2}} \nonumber \\ &&
 \left( \frac{v_0}{40~\mathrm{km\,s^{-1}}} \right) 
\left( \frac{B_\perp}{\mathrm{1 T}} \right)
 \mathrm{K},
\end{eqnarray}
where $B_\perp$ is the perpendicular component of the magnetic field and $y_\mathrm{t}$ is the number ratio of all the particles to hydrogen nuclei.
Since the magnetic field of accreting gas giants is highly uncertain, we are unable to validate our assumption of no magnetic effects at present. 
If $B_\perp \sim$ 1~T, according to the above estimates, the characteristic number density and temperature are comparable to those observed in the previous section, meaning the magnetic effects should be important for the post-shock processes. In reality, however, $B_\perp$ may be much smaller than 1~T. In any case, detailed investigate of the magnetic field of accreting gas giants is needed for resolving this issue.

\subsubsection{Effects of Preshock Heating}
\label{EPH}
We have performed numerical simulations only in postshock regions in this study, assuming the thermal
energy of the preshock gas is negligibly small relative to that of the postshock gas. 
However, absorption of radiative energy from postshock regions can heat the preshock gas and, thus, weaken the shock strength, because the Mach number becomes low. 
Consequently, the heating of the preshock gas leads to weakening the hydrogen line emission.

\citet{Marleau+2017} performed 1D radiative hydrodynamical simulations, taking account of radiative transfer in the preshock region. 
They obtained 2-3 times weaker hydrogen line luminosity than in the case without absorption. However, in contrast to this study, they assumed the local thermodynamic equilibrium, in which the electron energy states are uniquely determined at a given temperature, and simply the blackbody emission. 
This approximation is valid on large spatial scales that they were interested in. 
The scales of interest in this study are much smaller than theirs. 

On the other hand, \citet{Szulagyi+2016} performed 3D radiative hydrodynamical simulations and showed that the accreting gas giant could not have a circum-planetary disk, but have a circum-planetary envelope that extended to about the Hill radius.
Consequently, the shock strength at the surface of the circum-planetary envelope is too weak to excite the hydrogen atoms. 
However, as also shown in \citet{Szulagyi2017}, whether an accreting gas giant is surrounded by a disk or an envelope depends on the temperature of the protoplanetary surface, which is given as the numerical boundary condition. 
The hydrodynamic accretion simulations that also determine planetary temperature in a self-consistent fashion will be needed for clarifying the environments around accreting protoplanets.

\subsubsection{Effects of Thickness of Postshock Region}
We ended the numerical integration, once the gas temperature decreases to $1\times10^4$~K. 
At that point, the flow still retains about a half of its initial energy.
Thus, the actual value of line energy flux from the shock surface is up to twice as large as estimated above. 
However, in deeper regions we have ignored, molecules such as OH, CO, and H$_2$O make dominant contribution to cooling, instead of hydrogen line cooling, and have no significant influence on the radiative properties of the shallower regions.

While we consider only a vertical flow, there is also a flow rotating around a central protoplanet, namely a circum-planetary disk. The typical timescale on which both flows merge with each other is approximately the Keplerian period multiplied by the ratio of the vertical flow to horizontal flow densities. Since this timescale is longer than the time for which we have integrated, our assumption is valid.

\section{Summary and Conclusions}
\label{CS}
According to recent high-resolution 3D hydrodynamic simulations of accreting flow onto gas giants, the incoming gas falls vertically down to the surface of the circumplanetary disk and then passes through strong shockwaves. 
Because of strong shock heating, the gas becomes hot enough that hydrogen line emission is generated in postshock regions. To estimate the flux of the hydrogen line radiation, we have developed a 1D radiative hydrodynamic model of the flow after passing through the shockwave, performing the detailed calculations of
chemical reactions, electron transitions in hydrogen atoms, and radiative transfer. 

We have found that most of the energy that the flow has before shock is lost through radiative emission of hydrogen Lyman-$\alpha$ line.
Since the Lyman-$\alpha$ line is widely broadened by natural broadening and the postshock region is thin for the radiation from the line wing, absorption of the radiation from downstream by upstream gas has little influence on the Lyman-$\alpha$ flux at the shock front.
However, the absorption of line radiation has a great effect on the distribution of energy levels of electrons and enhances emission of other lines such as Balmer-$\alpha$ (H$\alpha$), Paschen-$\alpha$ and so on.

Integrating the energy flux throughout the surface of the circum-planetary disk, we have estimated the hydrogen line luminosity from an accreting gas giant as a function of protoplanet mass and circum-stellar disk gas density.
Then we have demonstrated that the H$\alpha$ luminosity could be strong enough as the source of the observed H$\alpha$ flux reported by \citet{Sallum+2015}, 
although the accretion process is to be examined in further detail for confirming whether the H$\alpha$ emission is of planetary origin.
Other lines in the atmospheric window such as Paschen-$\alpha$ and Paschen-$\beta$ could be observed with current observation instruments. 
Observation of hydrogen line emission from protoplanets is highly encouraged to obtain direct constraints to accreting gas giants, which will be key in understanding their formation.

\begin{figure*}[htbp]
\plottwo{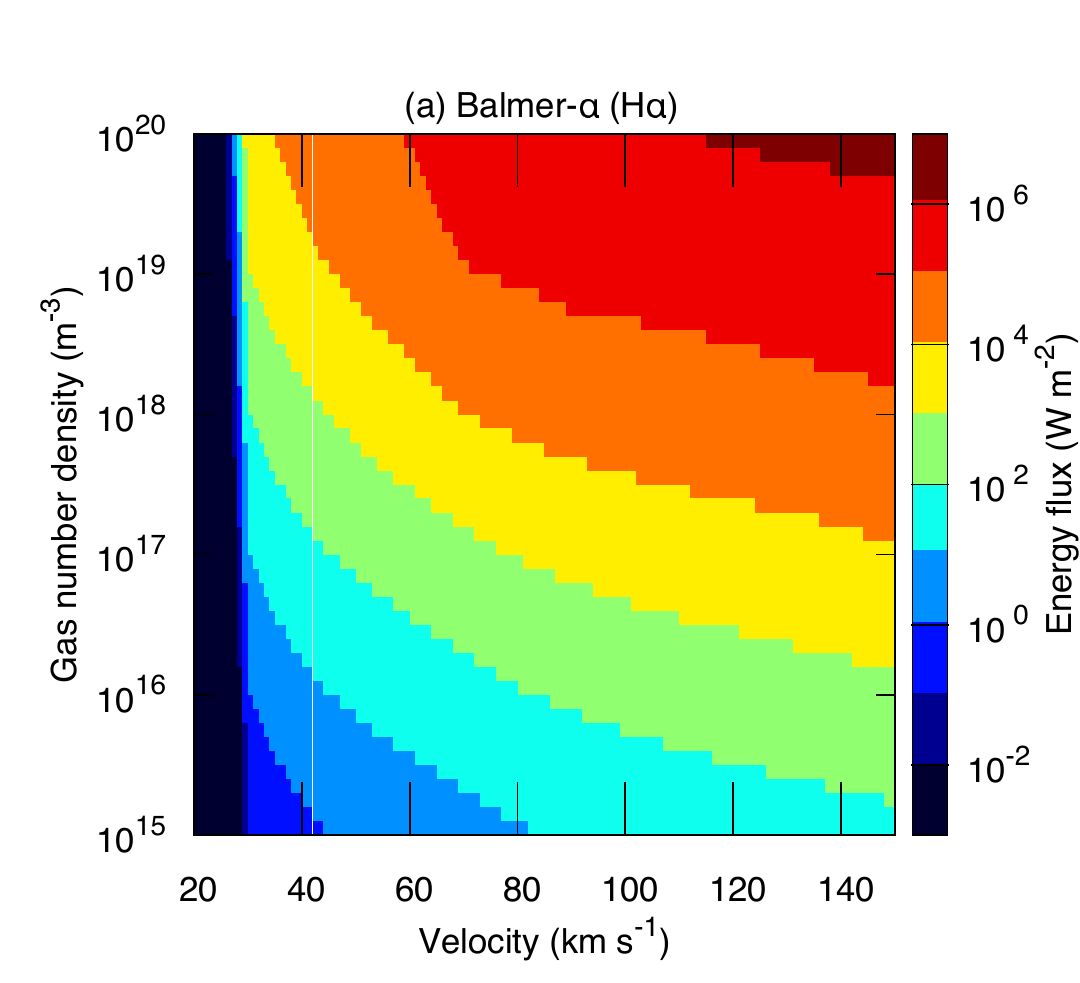}{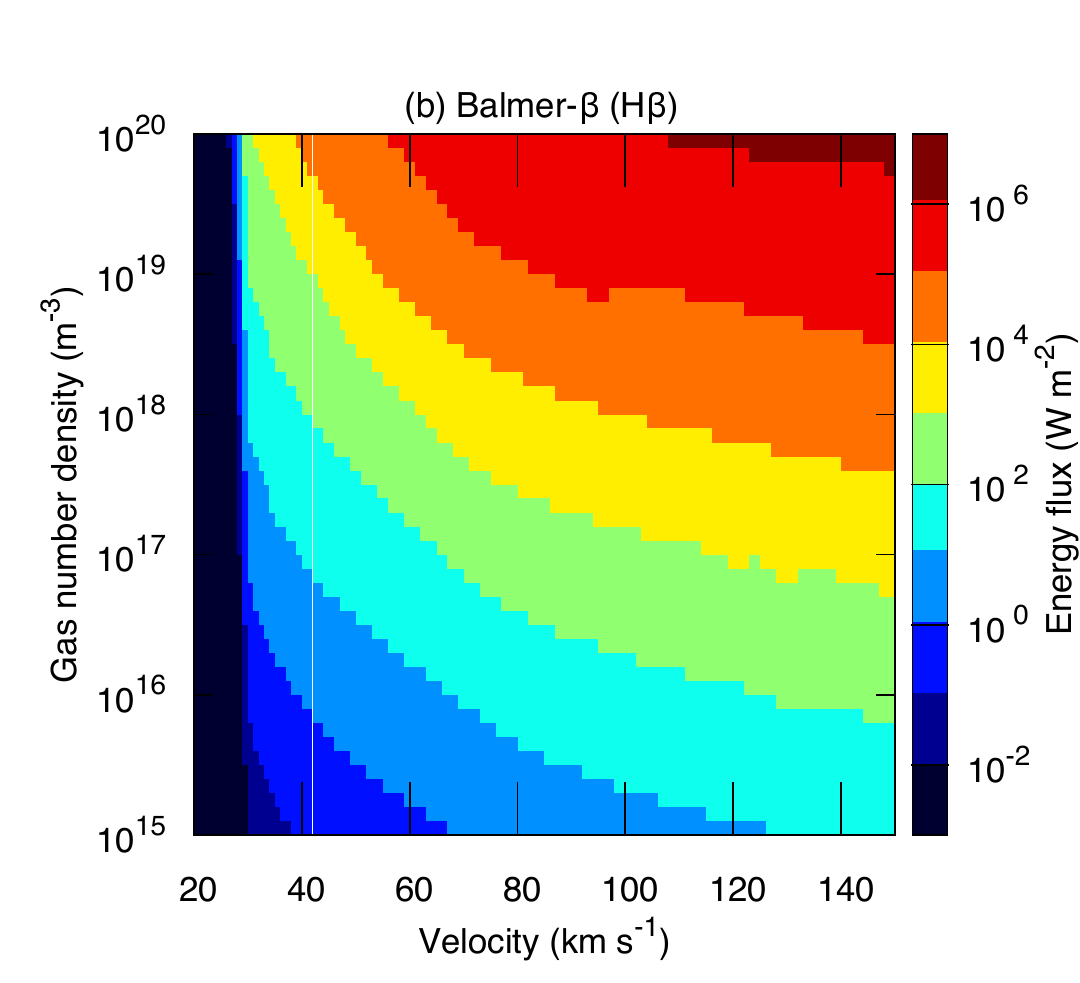}
\plottwo{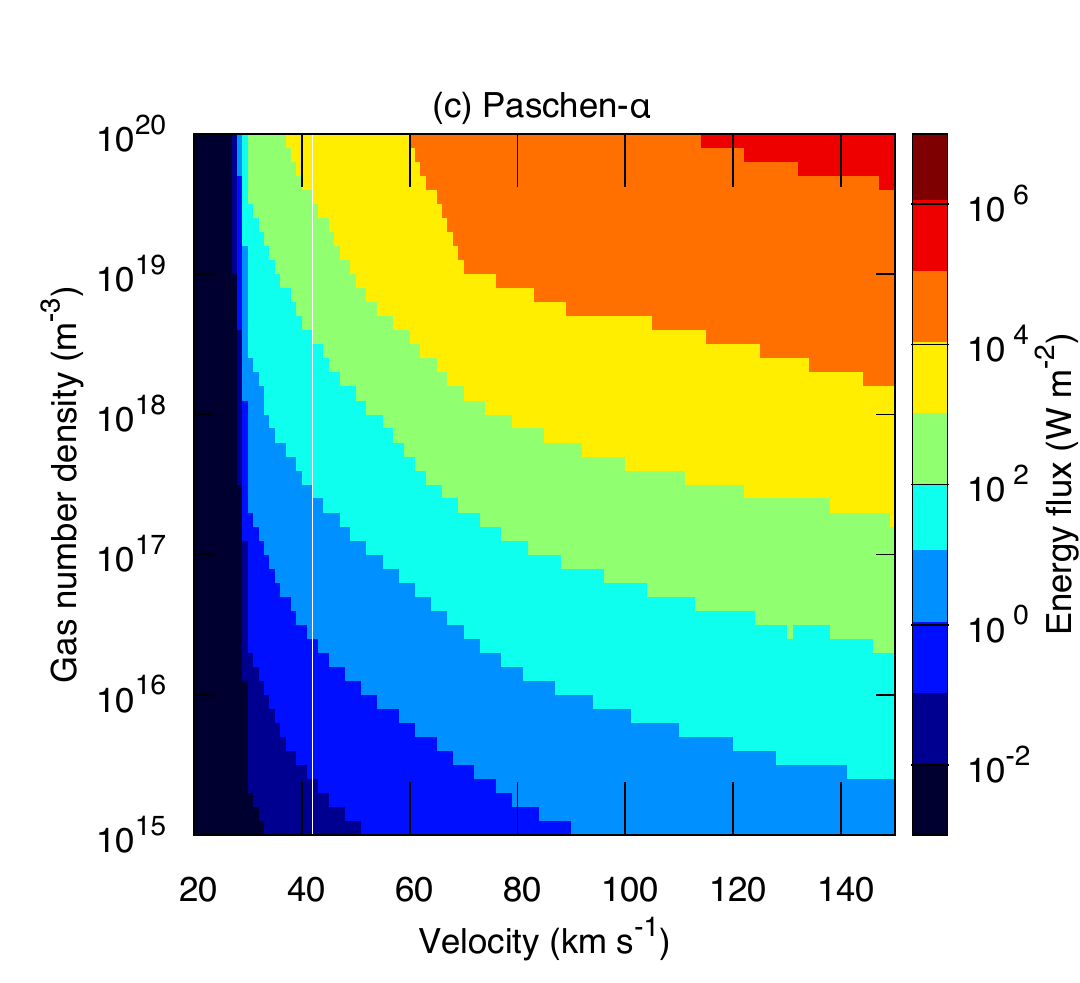}{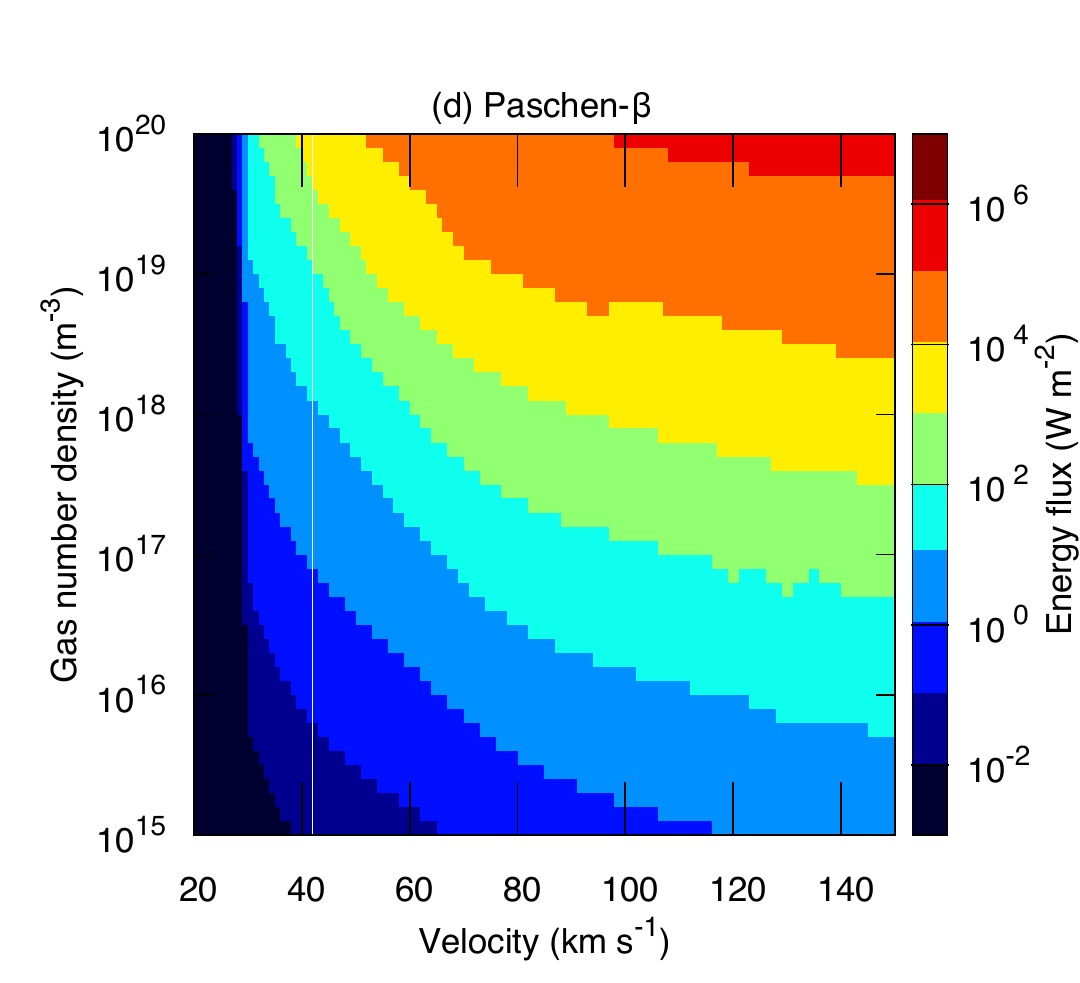}
\caption{
Contour plot of the energy flux of (a) Balmer-$\alpha$, (b) Balmer-$\beta$, (c) Paschen-$\alpha$, and (d) Paschen-$\beta$ lines versus the preshock velocity $v_0$ and total number density of hydrogen nuclei.
}
\label{fig:FC}
\end{figure*}

\appendix
Figure~\ref{fig:FC} is the color contour plot of the Balmer-$\alpha$ (H$\alpha$), Balmer-$\beta$, Paschen-$\alpha$, and Paschen-$\beta$ line energy fluxes versus the preshock velocity $v_0$ and the total number density of hydrogen nuclei $n_\mathrm{H,0}$. 

\acknowledgments
We would like to thank S. Inutsuka for his helpful suggestion about line profiles,
H. Kawahara for useful discussions about cooling and radiation processes,
and Y. Ito for his fruitful comments on the numerical scheme.
We thank the anonymous referee for his/her careful reading and constructive comments, which helped us improve this paper greatly.
This work was supported by JSPS KAKENHI Grant Numbers JP17H01153, JP18H05439, JP15H02065, and JP26800229. 
Y.A. was supported by Leading Graduate Course for Frontiers of Mathematical Sciences and Physics.
Y.A. and M.I. were also supported by JSPS Core-to-Core Program ``International Network of Planetary Science''.

\software{ODEPACK\citep{LSODE}}

\bibliographystyle{aasjournal}
\bibliography{./list}

\begin{thebibliography}{}
\expandafter\ifx\csname natexlab\endcsname\relax\def\natexlab#1{#1}\fi
\providecommand{\url}[1]{\href{#1}{#1}}
\providecommand{\dodoi}[1]{doi:~\href{http://doi.org/#1}{\nolinkurl{#1}}}
\providecommand{\doeprint}[1]{\href{http://ascl.net/#1}{\nolinkurl{http://ascl.net/#1}}}
\providecommand{\doarXiv}[1]{\href{https://arxiv.org/abs/#1}{\nolinkurl{https://arxiv.org/abs/#1}}}

\bibitem[{{Allen}(1976)}]{Allen3rd}
{Allen}, C.~W. 1976, {Astrophysical Quantities}

\bibitem[{{Biller} {et~al.}(2014){Biller}, {Males}, {Rodigas}, {Morzinski},
  {Close}, {Juh{\'a}sz}, {Follette}, {Lacour}, {Benisty}, {Sicilia-Aguilar},
  {Hinz}, {Weinberger}, {Henning}, {Pott}, {Bonnefoy}, \&
  {K{\"o}hler}}]{Biller+2014}
{Biller}, B.~A., {Males}, J., {Rodigas}, T., {et~al.} 2014, \apjl, 792, L22,
  \dodoi{10.1088/2041-8205/792/1/L22}

\bibitem[{{Blanksby} \& {Ellison}(2003)}]{Blanksby+Ellison2003}
{Blanksby}, S.~J., \& {Ellison}, G.~B. 2003, Chem. Res., 36(4), 255,
  \dodoi{10.1021/ar020230d}

\bibitem[{{Bodenheimer} \& {Pollack}(1986)}]{Bodenheimer+Pollack1986}
{Bodenheimer}, P., \& {Pollack}, J.~B. 1986, \icarus, 67, 391,
  \dodoi{10.1016/0019-1035(86)90122-3}

\bibitem[{{Calvet} {et~al.}(2002){Calvet}, {D'Alessio}, {Hartmann}, {Wilner},
  {Walsh}, \& {Sitko}}]{Calvet+2002}
{Calvet}, N., {D'Alessio}, P., {Hartmann}, L., {et~al.} 2002, \apj, 568, 1008,
  \dodoi{10.1086/339061}

\bibitem[{{Calvet} \& {Gullbring}(1998)}]{Calvet+Gullbring1998}
{Calvet}, N., \& {Gullbring}, E. 1998, \apj, 509, 802, \dodoi{10.1086/306527}

\bibitem[{{Castor}(2004)}]{Castor2004}
{Castor}, J.~I. 2004, {Radiation Hydrodynamics}, 368

\bibitem[{{Chandrasekhar}(1960)}]{Chandrasekhar1960}
{Chandrasekhar}, S. 1960, {Radiative transfer}

\bibitem[{{Frank} {et~al.}(1983){Frank}, {King}, \& {Lasota}}]{Frank+1983}
{Frank}, J., {King}, A.~R., \& {Lasota}, J.~P. 1983, \mnras, 202, 183,
  \dodoi{10.1093/mnras/202.1.183}

\bibitem[{{Goldreich} \& {Ward}(1973)}]{Goldreich+Ward1973}
{Goldreich}, P., \& {Ward}, W.~R. 1973, \apj, 183, 1051, \dodoi{10.1086/152291}

\bibitem[{{Gorti} {et~al.}(2011){Gorti}, {Hollenbach}, {Najita}, \&
  {Pascucci}}]{Gorti+2011}
{Gorti}, U., {Hollenbach}, D., {Najita}, J., \& {Pascucci}, I. 2011, \apj, 735,
  90, \dodoi{10.1088/0004-637X/735/2/90}

\bibitem[{{Gullbring} {et~al.}(1998){Gullbring}, {Hartmann}, {Brice{\~n}o}, \&
  {Calvet}}]{Gullbring+1998}
{Gullbring}, E., {Hartmann}, L., {Brice{\~n}o}, C., \& {Calvet}, N. 1998, \apj,
  492, 323, \dodoi{10.1086/305032}

\bibitem[{{Hayashi}(1981)}]{Hayashi1981}
{Hayashi}, C. 1981, Progress of Theoretical Physics Supplement, 70, 35,
  \dodoi{10.1143/PTPS.70.35}

\bibitem[{{Hern{\'a}ndez} {et~al.}(2008){Hern{\'a}ndez}, {Hartmann}, {Calvet},
  {Jeffries}, {Gutermuth}, {Muzerolle}, \& {Stauffer}}]{Hernandez+2008}
{Hern{\'a}ndez}, J., {Hartmann}, L., {Calvet}, N., {et~al.} 2008, \apj, 686,
  1195, \dodoi{10.1086/591224}

\bibitem[{Hindmarsh(2002)}]{LSODE}
Hindmarsh, A.~C. 2002, URL: http://www. llnl. gov/CASC/odepack [cited October
  18, 2005]

\bibitem[{{Hollenbach} \& {McKee}(1979)}]{HM79}
{Hollenbach}, D., \& {McKee}, C.~F. 1979, \apjs, 41, 555,
  \dodoi{10.1086/190631}

\bibitem[{{Hollenbach} \& {McKee}(1989)}]{HM89}
---. 1989, \apj, 342, 306, \dodoi{10.1086/167595}

\bibitem[{{Huml{\'{\i}}cek}(1982)}]{Humlicek1982}
{Huml{\'{\i}}cek}, J. 1982, \jqsrt, 27, 437,
  \dodoi{10.1016/0022-4073(82)90078-4}

\bibitem[{{Iida} {et~al.}(2001){Iida}, {Nakamoto}, {Susa}, \&
  {Nakagawa}}]{Iida+2001}
{Iida}, A., {Nakamoto}, T., {Susa}, H., \& {Nakagawa}, Y. 2001, \icarus, 153,
  430, \dodoi{10.1006/icar.2001.6682}

\bibitem[{{Ikoma} {et~al.}(2000){Ikoma}, {Nakazawa}, \& {Emori}}]{Ikoma+2000}
{Ikoma}, M., {Nakazawa}, K., \& {Emori}, H. 2000, \apj, 537, 1013,
  \dodoi{10.1086/309050}

\bibitem[{{Johns-Krull} {et~al.}(2016){Johns-Krull}, {McLane}, {Prato},
  {Crockett}, {Jaffe}, {Hartigan}, {Beichman}, {Mahmud}, {Chen}, {Skiff},
  {Cauley}, {Jones}, \& {Mace}}]{Johns-Krull+2016}
{Johns-Krull}, C.~M., {McLane}, J.~N., {Prato}, L., {et~al.} 2016, \apj, 826,
  206, \dodoi{10.3847/0004-637X/826/2/206}

\bibitem[{{Johnson}(1972)}]{Johnson1972}
{Johnson}, L.~C. 1972, \apj, 174, 227, \dodoi{10.1086/151486}

\bibitem[{{Koenigl}(1991)}]{Konigl1991}
{Koenigl}, A. 1991, \apjl, 370, L39, \dodoi{10.1086/185972}

\bibitem[{{Kraus} \& {Ireland}(2012)}]{Kraus&Ireland2012}
{Kraus}, A.~L., \& {Ireland}, M.~J. 2012, \apj, 745, 5,
  \dodoi{10.1088/0004-637X/745/1/5}

\bibitem[{{Lamzin}(1998)}]{Lamzin1998}
{Lamzin}, S.~A. 1998, Astronomy Reports, 42, 322.
\newblock \doarXiv{1303.4066}

\bibitem[{{Landau} \& {Lifshitz}(1959)}]{Landau+Lifshitz1959}
{Landau}, L.~D., \& {Lifshitz}, E.~M. 1959, {Fluid mechanics}

\bibitem[{{Lynden-Bell} \& {Pringle}(1974)}]{Lynden-Bell&Pringle1974}
{Lynden-Bell}, D., \& {Pringle}, J.~E. 1974, \mnras, 168, 603,
  \dodoi{10.1093/mnras/168.3.603}

\bibitem[{{Mac Low} \& {Shull}(1986)}]{MacLow+Shull1986}
{Mac Low}, M.-M., \& {Shull}, J.~M. 1986, \apj, 302, 585,
  \dodoi{10.1086/164017}

\bibitem[{{Marleau} {et~al.}(2017){Marleau}, {Klahr}, {Kuiper}, \&
  {Mordasini}}]{Marleau+2017}
{Marleau}, G.-D., {Klahr}, H., {Kuiper}, R., \& {Mordasini}, C. 2017, \apj,
  836, 221, \dodoi{10.3847/1538-4357/836/2/221}

\bibitem[{{Marley} {et~al.}(2007){Marley}, {Fortney}, {Hubickyj},
  {Bodenheimer}, \& {Lissauer}}]{Marley+2007}
{Marley}, M.~S., {Fortney}, J.~J., {Hubickyj}, O., {Bodenheimer}, P., \&
  {Lissauer}, J.~J. 2007, \apj, 655, 541, \dodoi{10.1086/509759}

\bibitem[{{Menu} {et~al.}(2014){Menu}, {van Boekel}, {Henning}, {Chandler},
  {Linz}, {Benisty}, {Lacour}, {Min}, {Waelkens}, {Andrews}, {Calvet},
  {Carpenter}, {Corder}, {Deller}, {Greaves}, {Harris}, {Isella}, {Kwon},
  {Lazio}, {Le Bouquin}, {M{\'e}nard}, {Mundy}, {P{\'e}rez}, {Ricci},
  {Sargent}, {Storm}, {Testi}, \& {Wilner}}]{Menu+2014}
{Menu}, J., {van Boekel}, R., {Henning}, T., {et~al.} 2014, \aap, 564, A93,
  \dodoi{10.1051/0004-6361/201322961}

\bibitem[{{Miki}(1982)}]{Miki1982}
{Miki}, S. 1982, Progress of Theoretical Physics, 67, 1053,
  \dodoi{10.1143/PTP.67.1053}

\bibitem[{Millikan \& White(1963)}]{Millikan+White1963}
Millikan, R.~C., \& White, D.~R. 1963, \jcp, 39, 3209,
  \dodoi{{10.1063/1.1734182}}

\bibitem[{{Mizuno}(1980)}]{Mizuno1980}
{Mizuno}, H. 1980, Progress of Theoretical Physics, 64, 544,
  \dodoi{10.1143/PTP.64.544}

\bibitem[{{Neufeld} \& {Kaufman}(1993)}]{Neufeld+Kaufman1993}
{Neufeld}, D.~A., \& {Kaufman}, M.~J. 1993, \apj, 418, 263,
  \dodoi{10.1086/173388}

\bibitem[{{Pollack} {et~al.}(1996){Pollack}, {Hubickyj}, {Bodenheimer},
  {Lissauer}, {Podolak}, \& {Greenzweig}}]{Pollack+1996}
{Pollack}, J.~B., {Hubickyj}, O., {Bodenheimer}, P., {et~al.} 1996, \icarus,
  124, 62, \dodoi{10.1006/icar.1996.0190}

\bibitem[{{Quanz} {et~al.}(2015){Quanz}, {Amara}, {Meyer}, {Girard},
  {Kenworthy}, \& {Kasper}}]{Quanz+2015}
{Quanz}, S.~P., {Amara}, A., {Meyer}, M.~R., {et~al.} 2015, \apj, 807, 64,
  \dodoi{10.1088/0004-637X/807/1/64}

\bibitem[{{Reggiani} {et~al.}(2014){Reggiani}, {Quanz}, {Meyer}, {Pueyo},
  {Absil}, {Amara}, {Anglada}, {Avenhaus}, {Girard}, {Carrasco Gonzalez},
  {Graham}, {Mawet}, {Meru}, {Milli}, {Osorio}, {Wolff}, \&
  {Torrelles}}]{Reggiani+2014}
{Reggiani}, M., {Quanz}, S.~P., {Meyer}, M.~R., {et~al.} 2014, \apjl, 792, L23,
  \dodoi{10.1088/2041-8205/792/1/L23}

\bibitem[{{Sallum} {et~al.}(2015){Sallum}, {Follette}, {Eisner}, {Close},
  {Hinz}, {Kratter}, {Males}, {Skemer}, {Macintosh}, {Tuthill}, {Bailey},
  {Defr{\`e}re}, {Morzinski}, {Rodigas}, {Spalding}, {Vaz}, \&
  {Weinberger}}]{Sallum+2015}
{Sallum}, S., {Follette}, K.~B., {Eisner}, J.~A., {et~al.} 2015, \nat, 527,
  342, \dodoi{10.1038/nature15761}

\bibitem[{{Seaton}(1959)}]{Seaton1959}
{Seaton}, M.~J. 1959, \mnras, 119, 81, \dodoi{10.1093/mnras/119.2.81}

\bibitem[{{Shapiro} \& {Kang}(1987)}]{Shapiro+Kang1987}
{Shapiro}, P.~R., \& {Kang}, H. 1987, \rmxaa, 14

\bibitem[{{Shu}(1991)}]{Shu1991}
{Shu}, F.~H. 1991, {The physics of astrophysics. Volume 1: Radiation.}

\bibitem[{{Szul{\'a}gyi}(2017)}]{Szulagyi2017}
{Szul{\'a}gyi}, J. 2017, \apj, 842, 103, \dodoi{10.3847/1538-4357/aa7515}

\bibitem[{{Szul{\'a}gyi} {et~al.}(2016){Szul{\'a}gyi}, {Masset}, {Lega},
  {Crida}, {Morbidelli}, \& {Guillot}}]{Szulagyi+2016}
{Szul{\'a}gyi}, J., {Masset}, F., {Lega}, E., {et~al.} 2016, \mnras, 460, 2853,
  \dodoi{10.1093/mnras/stw1160}

\bibitem[{{Szul{\'a}gyi} \& {Mordasini}(2017)}]{Szulagyi&Mordasini2017}
{Szul{\'a}gyi}, J., \& {Mordasini}, C. 2017, \mnras, 465, L64,
  \dodoi{10.1093/mnrasl/slw212}

\bibitem[{{Tanigawa} {et~al.}(2012){Tanigawa}, {Ohtsuki}, \&
  {Machida}}]{Tanigawa+2012}
{Tanigawa}, T., {Ohtsuki}, K., \& {Machida}, M.~N. 2012, \apj, 747, 47,
  \dodoi{10.1088/0004-637X/747/1/47}

\bibitem[{{Tanigawa} \& {Watanabe}(2002)}]{Tanigawa+2002}
{Tanigawa}, T., \& {Watanabe}, S.-i. 2002, \apj, 580, 506,
  \dodoi{10.1086/343069}

\bibitem[{{Thalmann} {et~al.}(2016){Thalmann}, {Janson}, {Garufi},
  {Boccaletti}, {Quanz}, {Sissa}, {Gratton}, {Salter}, {Benisty}, {Bonnefoy},
  {Chauvin}, {Daemgen}, {Desidera}, {Dominik}, {Engler}, {Feldt}, {Henning},
  {Lagrange}, {Langlois}, {Lannier}, {Le Coroller}, {Ligi}, {M{\'e}nard},
  {Mesa}, {Meyer}, {Mulders}, {Olofsson}, {Pinte}, {Schmid}, {Vigan}, \&
  {Zurlo}}]{Thalmann+2016}
{Thalmann}, C., {Janson}, M., {Garufi}, A., {et~al.} 2016, \apjl, 828, L17,
  \dodoi{10.3847/2041-8205/828/2/L17}

\bibitem[{{Uchida} \& {Shibata}(1984)}]{Uchida&Shibata1984}
{Uchida}, Y., \& {Shibata}, K. 1984, \pasj, 36, 105

\bibitem[{{Uyama} {et~al.}(2017){Uyama}, {Tanigawa}, {Hashimoto}, {Tamura},
  {Aoyama}, {Brandt}, \& {Ishizuka}}]{Uyama+2017}
{Uyama}, T., {Tanigawa}, T., {Hashimoto}, J., {et~al.} 2017, \aj, 154, 90,
  \dodoi{10.3847/1538-3881/aa816a}

\bibitem[{{van der Marel} {et~al.}(2015){van der Marel}, {van Dishoeck},
  {Bruderer}, {P{\'e}rez}, \& {Isella}}]{Marel+2015}
{van der Marel}, N., {van Dishoeck}, E.~F., {Bruderer}, S., {P{\'e}rez}, L., \&
  {Isella}, A. 2015, \aap, 579, A106, \dodoi{10.1051/0004-6361/201525658}

\bibitem[{{Vriens} \& {Smeets}(1980)}]{Vriens+Smeets1980}
{Vriens}, L., \& {Smeets}, A.~H.~M. 1980, \pra, 22, 940,
  \dodoi{10.1103/PhysRevA.22.940}

\bibitem[{{Zhu}(2015)}]{Zhu2015}
{Zhu}, Z. 2015, \apj, 799, 16, \dodoi{10.1088/0004-637X/799/1/16}

\end{thebibliography}
\end{document}